# OPTICAL AND INFRARED COLORS OF TRANSNEPTUNIAN OBJECTS OBSERVED WITH HST

S. D. Benecchi[1,2], K. S. Noll[3], D. C. Stephens[4], W. M. Grundy[5], J. Rawlins[4]

[1] Planetary Science Institute, 1700 East Fort Lowell, Suite 106, Tucson, AZ 85719
[2] Carnegie Institution of Washington, Department of Terrestrial Magnetism, 5241 Broad Branch Road, NW, Washington, DC 20015
[3] Space Telescope Science Institute, 3700 San Martin Dr., Baltimore MD 21218
[4] Brigham Young University, Dept. of Physics & Astronomy, N145 ESC, Provo, UT 84602
[5] Lowell Observatory, 1400 W. Mars Hill Rd., Flagstaff, AZ 86001

## ABSTRACT

We present optical colors of 72 transneptunian objects (TNOs), and infrared colors of 80 TNOs obtained with the WFPC2 and NICMOS instruments, respectively, on the Hubble Space Telescope (HST). Both optical and infrared colors are available for 32 objects that overlap between the datasets. This dataset adds an especially uniform, consistent and large contribution to the overall sample of colors, particularly in the infrared. The range of our measured colors is consistent with other colors reported in the literature at both optical and infrared wavelengths. We find generally good agreement for objects measured by both us and others; 88.1% have better than 2 sigma agreement. The median $H_V$ magnitude of our optical sample is 7.2, modestly smaller (~0.5 mag) than for previous samples. The median absolute magnitude, $H_V$, in our infrared sample is 6.7. We find no new correlations between color and dynamical properties (semi-major axis, eccentricity, inclination and perihelion). We do find that colors of Classical objects with $i<6°$ come from a different distribution than either the Resonant or excited populations in the visible at the >99.99% level with a K-S test. The same conclusion is found in the infrared at a slightly lower significance level, 99.72%. Two Haumea collision fragments with strong near infrared ice bands are easily identified with broad HST infrared filters and point to an efficient search strategy for identifying more such objects. We find evidence for variability in (19255) 1999 $VK_8$, 1999 $OE_4$, 2000 $CE_{105}$, 1998 $KG_{62}$ and 1998 $WX_{31}$.

# 1.    INTRODUCTION

## *1.1.    Transneptunian Colors*

Small body populations in the solar system are remnants of the earliest phases of accretion from the protoplanetary disk. Their physical properties are a function of their formation and subsequent migration and evolution. In the Kuiper Belt, there is evidence that the wide range of optical colors is related to primordial compositional differences in these objects (Tegler et al. 2003, Benecchi et al. 2009).  Links between the dynamical and photometric properties of these objects may, therefore, help to distinguish between various source populations and the range of conditions present in the protoplanetary disk.

Colors derived from filter photometry were among the first things known about transneptunian objects (TNOs). Early studies by Jewitt & Luu (2001) made it clear that these objects had an unusually large range of colors. This result fueled a number of efforts to search for correlations between color and the dynamical state of TNOs including this work. Due to the faintness of TNOs, the majority of photometric work (until recently) was focused in optical wavelengths. The only strong correlation found is for objects residing in the low inclination, low eccentricity, 'cold' Classical portion of the belt. They are, on average, redder at optical wavelengths (Tegler & Romanishin 1998; Peixinho et al. 2004; Gulbis et al. 2006; Peixinho et al. 2008) than objects at higher inclination in otherwise similar orbits. The Centaur population, scattered objects with orbits between Jupiter and Neptune, has a color distribution that appears to be bimodal (Peixinho et al. 2003; Tegler et al. 2008) while the Scattered, Plutino (3:2) and Classical (*i<12°*) populations show no particular trends (Doressoundiram et al. 2008; Romanishin et al. 2010). Studies of objects at the extremes of the Kuiper Belt (beyond the edge of the Classical region: high inclination, large semi-major axes or large perihelion distances) show them to be very red (Sheppard 2010), although samples for individual extreme groups are small. For all these studies, the results are strongest in the B-R color, but are seen throughout the optical range.

In this paper we present photometric measurements of TNOs at optical and infrared wavelengths resulting from two Hubble Space Telescope (HST) programs executed in 2001—2003. The objects observed were a subset of the full target list included as potential targets for these two "snapshot" programs. Potential targets were selected based on predicted positional uncertainty and expected target brightness.

## *1.2.    Nomenclature*

In this paper we adopt a modified dynamical classification scheme using information available online (http://www.boulder.swri.edu/~buie/kbo/astrom/) or in the literature: Elliot et al. (2005, DES) and Gladman et al. (2009, GMV). We sort objects into one of three possible categories: Resonant, Cold, or Hot. Resonant objects are defined identically in the classification schemes of both DES and GMV, they include any object in a mean motion resonance with Neptune. Our Cold group includes Classicals with *i≤6°*, another area of agreement between DES and GMV. Our Hot grouping includes objects defined as Classicals with *i>6°*,

Scattered/Scattering, Detached, and Centaurs. The definitions of these subgroups are where the DES and GMV differ. However, by grouping these objects together into a "Hot" category, the two classifications are again in agreement. There is abundant physical evidence for separating low inclination Classicals as a distinct group including their color, as noted above, their high concentration of binaries (Noll et al. 2008) and their high albedos (Brucker et al. 2009). According to giant planet migration models (e.g. the *Nice model*, Morbidelli et al., 2005, Levison et al., 2008, Tsiganis et al., 2005, Gomes et al., 2005), much of the rest of the transneptunian population may be a mix of planetesimals scattered from a broad region of the protoplanetary nebula. There is no current unambiguous physical evidence of distinct subpopulations to contradict this view. Indeed, it might well make sense to consider Resonant objects as part of the Hot population as well, although we retain this distinction here. About half our sample are Cold objects, with one quarter of the sample each in the Resonant and Hot groupings. We do not measure enough objects in any single Resonant population to warrant dividing the objects farther. Likewise, we only measure a few Centaurs so we are unable to separate out this population for additional study.

## 2. OBSERVATIONS

In this paper we report observations associated with two different programs executed in HST Cycles 10 (June 2001 – July 2002) and 11 (August 2002 – June 2003).

The first program, proposal 9060, observed 72 TNOs using the Wide Field Planetary Camera 2 (WFPC2) in three broad filters: F555W (0.5442 μm) ~ V, F675W (0.6717 μm) ~ R, and F814W (0.7996 μm) ~ I. Objects were placed on the Wide Field Camera 3 detector which has a pixel scale of 0.1 arcsec/pixel. The objects were selected based on their small orbital uncertainties (≤30 arcseconds), without respect to dynamical classification, and had expected V magnitudes between 22 and 24. Objects predicted to be brighter than $m_v$=23.5 followed the observing sequence: 1x160 seconds in F555W, 2x160 seconds in F675W, 2x160 seconds in F814W, 1x160 seconds in F555W. For fainter objects, the F675W measurement was excluded and the exposure time was increased to 260 seconds for the other filters. The second program, proposal 9386, observed 80 TNOs using the Near Infrared Camera and Multi-Object Spectrometer (NICMOS), in two filters: F110W (1.1 μm) ~ J and F160W (1.6 μm) ~ H. The observing sequence was: 1x256 seconds in F110W, 2x512 seconds in F160W, 1x256 seconds in F110W. Objects were observed with the NIC2 camera which has a pixel scale 0.075 arcsec/pixel.

Observations in the WFPC2 program were not dithered. Observations with Nicmos were dithered to mitigate the impact of hot pixels and other fixed pattern noise. The observations were bracketed by the F555W filter in the optical, and the F110W filter in the infrared, to look for lightcurve effects; no significant variations were found on orbit timescales (~50 minutes) in the optical data (Stephens et al. 2003) nor in the infrared. Table 1 lists objects from both programs, their orbital properties and observational circumstances. Two dynamical classifications are listed, one as discussed in section 1.2 (column 3) and the DES classification (column 4).

INSERT TABLE 1 HERE

## 3. HST CALIBRATION AND PHOTOMETRY

Data were processed through the standard HST pipelines for WFPC2 and NICMOS

(Pavlovsky et al. 2006; Dressel et al. 2007). The pipeline flags static bad pixels, performs A/D conversion, subtracts the bias, and dark images, and corrects for flat fielding. It also updates the header with the appropriate photometry keywords.

WFPC2 data calibration is further described in Stephens et al. (2003) and includes the additional steps of geometric correction, removal of the '34th row effect'[1] (Anderson & King 1999) and a charge transfer efficiency (CTE) correction. NICMOS data calibration included two additional steps: (1) correction for quadrant dependent bias variations using the STSDAS software tool PEDSKY (iterative mode; Thatte & Dahlen et al. 2009) and (2) correction for rate-dependent nonlinearity on a pixel-by-pixel basis using a corrected version of the RNLINCOR tool that properly treats the background (de Jong 2006) and is necessary for accurate photometry, particularly for faint sources such as we observed.

For both WFPC2 and NICMOS data, photometry was extracted by fitting the observed Point Spread Function (PSF) with a model PSF generated by the TinyTim software tool (Krist & Hook 2004). The methodology we used follows the same procedure that was described in Stephens & Noll (2006). Because the PSF varies with location and color of the object, we generated a polychromatic PSF at the pixel location of each individual observation using preliminary measurements of the object's color. In order to allow for subpixel centroiding, we oversampled the PSF by a factor of 10 in both the x and y (row and column measured in pixels) dimensions. We then rebinned this model to the dimensions of the image, but adjusted the rows and columns during the binning process so that we could iteratively shift the center of the PSF by $1/10^{th}$ of a pixel in the x and y dimensions.

An initial value for the scaling factor, z (in counts), was determined by summing the counts in the pixels of the actual image. Holding this initial estimate of the scale fixed, we shifted the model PSF in the x and y directions until the best center was found. We then varied the z value in increments of $1/1000^{th}$ of a count, until we found the scaling factor that minimized the overall image residual. Using the new estimate of the scale, z, we again adjusted the x and y coordinates of the PSF until the best center was found, and then adjusted the scaling factor z for the new center. This process was iteratively repeated until we found the x, y, z combination that produced the smallest residual.

The residuals were calculated by subtracting the model PSF from the actual image and taking the sum of the squares of a group of pixels centered around the object. In calculating the residuals, we only used those pixels where the counts in the original image were at least one-sigma above the background sky. For the faintest objects, this roughly corresponded to a 3x3 box of pixels that were used in the calculation. For the brightest objects, this corresponded to a 5x5 box of pixels. The model PSF with the smallest residual determined the best-fit centroid for the object $x_o$, $y_o$ and the scaling factor, $z_o$. The approach described above was chosen over traditional 2-D Gaussian PSF fitting because the TNOs are faint and undersampled. We modeled our approach after a paper written by Howell et al. 1996, where he shows that such a technique is best for determining point-source centroids and brightnesses for faint objects with high precision.

Photometry for each object was calculated from the TinyTim PSF with the best fit scaling factor, $z_o$, using standard aperture photometry. For the WFPC2 data, the aperture size, CTE correction, and zeropoints were taken from values provided by Dolphin (2000; 2009). This produced absolute magnitudes on the Vega system. For the NICMOS data, the counts in an aperture of 6.67 pixels were measured and then corrected to an absolute magnitude using the most recent set of photometric keywords, Vegamag zeropoints, and aperture corrections. These keywords and corrections are available on the NICMOS webpage as described in the NICMOS Data Handbook (Thatte, D. & Dahlen, T. et al. 2009; hard copies of the webpages at the time of this paper can be found in online supplemental material[2]).

---

[1] A manufacturing defect in the WFPC2 CCD chip such that every $34^{th}$ row is approximately 3% narrower than all others. It induces errors in the photometry at the 0.01-0.2 magnitude level if not corrected for following the methodology provided by Anderson & King 1999).

To determine the error in the final magnitudes for each object, we used the residuals that were calculated for the various combinations of x, y, and z during the initial fitting process. We wanted to determine which of the other model PSFs could be considered equally probable of being the correct fit to the data within a one-sigma error. To determine which models to include, we took the square-root of the reduced chi-squared value of the best fit model $x_o$, $y_o$, and $z_o$. We added this value to the reduced chi-squared fit of the best model and considered all model PSFs that had reduced chi-squared values less than this total amount to be equally good fits to the data. We then calculated the magnitudes for all of these models and used the upper and lower extremes in magnitude to define the error. The magnitudes and their uncertainties are tabulated in Table 2.

INSERT TABLE 2 HERE

Once photometry for each object was determined in the HST filters we converted the magnitudes to standard Johnson-Cousins-Bessel filters (F555W —> Johnson V; F675W —> Cousins R; F814W —> Cousins I; F110W —> Bessel J; and F160W —> Bessel H) for comparison with other datasets in the literature. The conversion factors are different for each object, camera and filter. For each object we take its color in the HST filters, F555W-814W and F110W-F160W, and use *synphot* with a solar spectrum (Kurucz model) reddened to match that of the object (in the respective filters) to determine the appropriate offsets. We add the offset values to the magnitudes for each object in the HST filter system to find the nominal magnitude of the object in standard filters. To determine the uncertainties on the converted magnitudes, we consider the 1 sigma minimum and maximum colors in the HST filters (color plus and minus 1 sigma error bar) for a given object and determine the respective 1 sigma minimum and maximum offsets to the standard filters as described for the nominal magnitude values. A sample correction calculation can be found in Appendix A, paragraph 3 of Benecchi et al. (2009). The resultant ground-based photometry values from these calculations are listed in Table 3.

INSERT TABLE 3 HERE

## 4. RESULTS

### 4.1. Photometry

Photometry for 120 TNOs is listed in Table 2. No previous photometry in the measured bandpasses has been reported for approximately half of these objects. Objects are listed in order of their permanent number and in order of their provisional id for those not yet numbered. Optical and/or infrared results are listed with 32 of the objects having both optical and infrared

---

measurements.

Photometry is reported in the Vega magnitude system. Uncertainties are minimized by reporting the magnitudes in the native HST filter bandpasses, as we do in Table 2. Objects ranged in magnitude from $21.90 \leq m_{F555W} \leq 24.97$ with a median magnitude of 23.64. In general, objects were found to be fainter than predicted by ~0.24 magnitudes. We also report our photometric measurements in the more frequently used Johnson-Cousins-Bessell filters in Table 3. The details of converting to these filters are described in Section 3.1.

Errors are determined as described in section 3.1. Typical best-case uncertainties are +/- 0.04 mag with WFPC2 and +/- 0.03 mag with NICMOS. These approach the expected performance limits for absolute photometry with both instruments (Baggett et al. 2002; McMaster et al. 2008 and Thatte et al. 2009; Viana et al. 2009, respectively). In some cases, the uncertainties are larger due to a combination of cosmic rays, compromised pixels and other detector artifacts, faint background sources, or possible intrinsic variability. There may also be cases where we just have a lower signal to noise observation than expected because the background values were higher (due to observations taken at different times of the year, the position of our observations relative to the passage of HST through the SAA or other background variation).

### 4.2. Absolute Magnitudes

We calculate absolute magnitudes at zero phase in the Johnson V band, $H_v$, for the 72 objects with measured F555W magnitudes. We compute the flux that would be observed at zero phase angle by using the linear phase function approximation:

$$H_V = m_{F555W \to V} - 5 \log R_{sun} \Delta - \beta \alpha \qquad \text{(Eq. 1)}$$

The geometric circumstances (where $R_{sun}$ is the heliocentric distance, $\Delta$ is the geocentric distance and $\alpha$ is the phase angle) are given in Table 1 and we use $\beta$=0.15 mag/° (Sheppard & Jewitt, 2002). The phase correction is applied to the measured magnitudes, after converting the HST F555W magnitude to the Johnson V magnitude ($m_{F555W \to V}$), to calculate $H_v$. The resulting values are tabulated in Table 4. Comparing our values with those of the Minor Planet Center (MPC; http://www.minorplanetcenter.org/iau/lists/Centaurs.html and TNOs.html on 9 October 2010) we find differences of up to 1.26 magnitudes with a median difference of 0.28 magnitudes. Likewise, comparing the 26 objects we have in common with those measured by Tegler and Romanishin (http://www.physics.nau.edu/~tegler/research/survey.htm), we find differences of up to 0.65 magnitudes with a median difference of 0.08 magnitudes. We note that Romanishin & Tegler (2005) previously reported a similar disagreement relative to the absolute magnitudes tabulated by the MPC.

INSERT TABLE 4 HERE

It is worth noting that Sheppard (2007) has suggested that main belt asteroid type phase curves (Bowell et al. 1989), as utilized by Tegler and Romanishin and the MPC, are not necessarily appropriate for TNOs. Santos-Sanz et al. (2009) calculate both $H_{V,Bowell}$ and $H_{V,Linear}$ magnitudes for their 32 object sample and find the Bowell calculation to yield a magnitude 0.04-0.06 brighter than the linear approximation (as used here) for typical TNO phase angles of 0.1-1.2°. The different methods used to correct to zero phase cannot, however, account for the significant difference between the absolute magnitudes we derive and those reported by the MPC. The agreement between our results and those of the MPC and Tegler & Romanishin would be slightly better (a median difference of 0.24 and 0.07 magnitudes respectively) if the phase correction were to be calculated based on the linear approximation for these two datasets.

We also note that the HST filters are not a perfect match to the ground based filter set, they are typically broader and include spectral features not seen in ground based bandpasses, in particularly in the I and J regions (F814W peaks much shorter than Cousins I and F110W is much wider than the Bessel J band). Transformations between filter sets introduce unavoidable uncertainties. We include formal uncertainties in our error bars, but systematic effects (e.g. differences in the unknown actual spectrum of an object compared to the one used in the transformation) may account for some of the imperfect agreement between our sample and others.

An important component of our data is the sensitivity to much smaller objects than those in other samples or than those observable from the ground in comparable observing time. The median absolute magnitude, $H_v$, of our optical and infrared samples are 7.2 and 6.7 which corresponds to diameters of ~150 and ~190 km assuming an albedo of 0.10. The smallest object in our sample is ~50 km in diameter with the same assumptions, a size where collisional evolution may be important (Pan & Sari 2005). The median diameter for objects observed with WFPC2 is ~50 km smaller than the median diameter in other samples, ~200 km (using the same albedo assumption). While we recognize that some radiation processing mechanisms can work rapidly, we suggest that objects in this small size range are dominated by erosion and that the colors, therefore, are potentially more directly related to primordial composition.

### *4.3.   Colors*

In Figures 1-4 we plot the colors derived from our photometry in Tables 2 and 3. Objects in the Cold, Hot and Resonant dynamical groupings are denoted by orange squares, blue diamonds, and black triangles, respectively.

INSERT FIGURE 1, FIGURE 2, FIGURE 3 AND FIGURE 4 HERE

As shown in Figures 1 and 2, optical colors range from slightly bluer than the Sun (for (130391) 2000 JG$_{81}$, F555W-F814W = 0.83) to significantly redder (1998 XY$_{95}$, 1993 SC and 1996 TQ$_{66}$, F555W-F814W = 1.54-1.67).

At infrared wavelengths the colors similarly range from slightly bluer to red relative to the Sun, with two exceptions. (86047) 1999 OY$_3$ and (24835) 1995 SM$_{55}$ are > 0.6 magnitudes bluer than the Sun in F110W-F160W (Figures 2 and 3). Both have been identified as probable members of the Haumea collision family (Ragozzine & Brown 2007). These objects are characterized by extremely strong near-infrared water ice bands (Snodgrass et al. 2010). We note that these objects are easily separated from the rest of the measured TNOs using the broad F110W and F160W filters (Noll et al. 2005). Narrower filters are not needed to identify these objects with HST.

Additionally, we note that ten of our targets: (86047) 1999 OY$_3$, (24835) 1995 SM$_{55}$, 1996 RQ$_{20}$, 1999 OH$_4$, 2001 QC$_{298}$, 2000 CG$_{105}$, and 1999 CD$_{158}$, (86177) 1999 RY$_{215}$, (130391) 2000 JG$_{81}$, and (20161) 1996 TR$_{66}$, were considered potential Haumea family candidates or diffused TNOs nearby based on dynamical studies in Ragozzine & Brown (2007). In their search of the photometric literature and in subsequent studies by Snodgrass et al. (2010) all but the first two objects were discounted from family membership due to their lack of infrared color agreement with Haumea. Likewise, we find these objects to have extremely different colors from Haumea in the infrared. However, we find 2001 QC$_{298}$, 2000CG$_{105}$, and (130391) 2000 JG$_{81}$ to be blue in the optical (V-I=1.00±0.08, 1.11±0.13, and 0.80±0.12), the last being the bluest object in our sample and the first also having neutral infrared colors (F110W-F160W=0.61±0.18). Snodgrass et al. (2010) argue that these objects are not family members, we suggest that perhaps these are fragments that were contaminated by less blue, non-ice material during disruption of the parent body or during subsequent re-accretion of the disrupted material. Cook et al. (2011) suggest that the low mass and low ejection speeds of icy Haumea family members suggests that there may be a similar mass in partially differentiated bodies that lack the strong spectral features of pure ice. Water ice is perhaps a sufficient but not necessary requirement for family membership.

We calculated the spectral gradient for each of our objects following the method of Hainaut & Delsanti (2002) and Boehnhardt et al. (2002). We defined the gradient relative to the F555W filter (our defined reference) as:

$$S(F555W, F814W) = 10^4 \left[ \frac{R(F814W) - R(F555W)}{\lambda_{F814W} - \lambda_{F555W}} \right], \qquad \text{(Eq. 2)}$$

where R(F555W) and R(F814W) are the normalized reflectances,

$$R(\lambda) = 10^{-0.4\left[\left(m_\lambda - m_{ref}\right) - \left(m_{\lambda,sun} - m_{ref,sun}\right)\right]}, \qquad \text{(Eq. 3)}$$

and $m_\lambda = m_{F555W}$ and $m_{F814W}$, respectively and $m_{\lambda,sun}$ and $m_{ref,sun}$ are given in Table 2. Our values are tabulated in the final column of Table 2. Averaging the values with respect to dynamical class we find gradients of: 9.35±4.55 for the cold objects ($i < 6°$), 8.18±5.89 for the all the resonant objects combined, and 6.79±4.72 for all the hot objects combined. Within the error bars, there appears to be no great difference between the dynamical classes. Our results are not inconsistent with those of other studies (Santos-Sanz et al. 2009; Hainaut & Delsanti 2002; and Boehnhardt et al. 2002, 2003), however, we note that our dataset lacks measurements in the "B" (or Blue, $\lambda \sim 4372$ angstroms) filter. Some studies suggest that TNO colors inclusive of the B filter present a different picture than that of solely VRI colors (Doressoundiram et al. 2007, Stephens et al. 2003). It may be that there is an active component on TNO surfaces that is only observable in the B spectral region (Doressoundiram et al. 2005; Santos-Sanz et al. 2009).

### 4.4. Comparisons

For more direct comparison with ground-based observations we convert our HST colors to standard Johnson-Cousins VRI and Bessel JH filters. We note that the literature is a rather inhomogeneous sample and recognize the work of Hainaut & Delsanti (2002) in their generation of the MBOSS (Minor Bodies of the Solar System) database. However, in order to be able to reference primary sources and to be confident of the values we are using, we generate our own database for comparison to measurements in the literature. Our comparison sample is based on the following criteria: (1) the dataset has reported VRI or JH colors, (2) each set of colors was taken during the same epoch, i.e. VRI or JH measurements were taken within the same few hours, (3) the photometric calibration was done with Landolt standard stars or other stars calibrated to the Landolt filter system (Johnson BV, Cousins RI), and (4) there were ~20 or more objects in the sample. We note that it is imperative that colors be collected as close in time to each other to minimize lightcurve complications and that the filter network employed for photometric calibration among different datasets is identical.

Published observations include measurements collected in Bessel, Johnson-Cousins and Mould filters. Each one of these filters has a slightly different response at the wavelengths of interest and in particular in the R and I bandpasses. Knowing which filter set is used is critical for proper combination of different datasets unless the same standard star network (or at least the same filter set) is referenced by all. Bessel filters have effectively the same bandpasses in the R and I wavelengths as Cousins filters, however the Johnson V-R solar color (referenced in Cox 2000) is 0.2 magnitudes different than the Cousins V-R solar color and Mould colors are different still. By using the Landolt standard star network, or other stars calibrated using this network/filter set, for absolute calibration, all colors are being normalized to Johnson V, Cousins R and Cousins I regardless of which filter system is physically being used by the instruments. Filter systems in the infrared are Bessel and calibrated with the UKIRT standard star system (Hunt et al. 1998; Persson et al. 1998; Hawarden et al. 2001).

For objects with multiple measurements, colors were combined and weighted with respect to their reported errors using the same formulation as Hainaut & Delsanti (2002):

$$\bar{x} = \frac{\displaystyle\sum_{i=1}^{N} \frac{x_i}{\sigma_i}}{\displaystyle\sum_{i=1}^{N} \frac{1}{\sigma_i}}, \text{ and } \sigma = \sqrt{\frac{\displaystyle\sum_{i=1}^{N} \sigma_i}{\displaystyle\sum_{i=1}^{N} \frac{1}{\sigma_i}} + \frac{\displaystyle\sum_{i=1}^{N} \frac{\left(x_i - \bar{x}\right)^2}{\sigma_i}}{(N-1)\displaystyle\sum_{i=1}^{N} \frac{1}{\sigma_i}}} \ , \qquad (Eq. 4)$$

where x is the color converted into flux space and σ is the uncertainty on the color converted into flux space. Carrying out these calculations in flux space is essential. If the values differ by any more than a typical color uncertainty, 0.1 magnitudes, the difference between averaging in flux vs. magnitude space can be as large as a few hundredths of a magnitude and the uncertainties will be overestimated.

Sixty-one objects have colors that can be compared to measurements in the literature. Of the 126 individual color measurements (V-R, R-I, V-I and J-H) where our measurements overlap with published colors, we find agreement within one sigma for 68.3% of our value and within 2 sigma for 91.3% of our values. The largest disagreement is in the 42 V-I color measurements where only 81.0% of objects have colors within 2 sigma of our values. All of the J-H colors are consistent within the error bars. It is worthwhile noting that the F814W and Cousins I bandpasses are not well matched, perhaps we are picking up signal in the F814W that is not strong in data from ground based observatories. It is also possible that objects with large differences between measurements are indicative of lightcurves, or other interesting surface/geometric effects. The objects of greatest interest include: (38083) 1998 HB$_{12}$, 1996 TK$_{66}$, (91133) 1998 HK$_{151}$, 1999 OE$_4$, 1998 WS$_{31}$, 1998 UR$_{43}$, 2000 CG$_{105}$, and (138537) 2000 OK$_{67}$. Of this list only (91133) 1998 HK$_{151}$ has any published variability data, suggesting variation at the level of <0.15 magnitudes (Sheppard & Jewitt, 2002). Attributing these offsets to surface effects would likely be hard to disentangle given the datasets available.

In Figure 4 we plot TNO colors from literature sources (see caption for reference details) in addition to our converted HST results. As expected our results overlap quite well. We looked for correlations, using a Spearman rank test, between object color separated by dynamical grouping (Cold, Hot and Resonant) and semi-major axis, eccentricity, inclination and perihelion. Using the HST dataset alone we don't find any statistically significant correlations. Combining the HST dataset with the literature we have a sample of 299 objects and find a >3 sigma correlation between the Hot population and inclination in the optical; the correlation in the infrared is not as strong ~2 sigma). We find a >3 sigma correlation between the Resonant, Hot and Centaur populations with absolute magnitude in the infrared. In the optical, the Hot-Centaur population yields a ~2 sigma result while the Resonant and Centaur populations have very little correlation (Table 5, values of interest are highlighted).

Additionally using a Kolmogorov-Smirnov (K-S) test we consider, for each of the colors and dynamical groups, the hypothesis that the groups come from the same parent population for both the HST values alone and then for the HST and literature values combined. The results of our calculations can be found in Table 6. We conclude, similar to other literature findings (Santos-Sanz et al. 2009) that the Cold population very likely came from a different distribution than either the Hot or Resonant populations. The results of the HST dataset alone are reasonably strong (92.4%), but the results including the literature find a better than 99.999% probability that these populations are distinct (inclusive or exclusive of the Centaurs in the Hot population). We draw the same conclusion from the infrared dataset, but at a lower significance level (88.00% and 99.72% in the HST only and combined datasets, respectively).

INSERT TABLE 5 AND TABLE 6 HERE

### 4.5. Optical and Infrared color overlap

Thirty-two objects in our sample have measurements in 4 or 5 HST filters. An additional 5 objects have infrared colors in our dataset and HST optical colors from Benecchi et al. 2009. In Figures 5 and 6 we plot normalized reflectance (to F555W) vs. wavelength for each object following the example of Hainaut & Delsanti (2002):

$$R_{\lambda,\lambda R} = 10^{-0.4\left[\left(m_\lambda - m_{\lambda R}\right) - \left(m_{\lambda,sun} - m_{\lambda R,sun}\right)\right]},$$ (Eq. 5)

where $m_\lambda$ is the magnitude of the object at wavelength $\lambda$, $m_{\lambda,sun}$ is the magnitude of the Sun at wavelength $\lambda$. $m_{\lambda R}$ and $m_{\lambda R,sun}$ are the magnitude of the object and the sun in the reference filter, F555W, respectively. The solar colors we used can be found in Table 2, they are based on the solar magnitudes and colors from Cox (2000) and converted using *synphot*. Each object is offset by 1 on the vertical scale for clarity.

INSERT FIGURE 5 AND FIGURE 6 HERE

Because the infrared data were collected at a different epoch than the optical we are susceptible to systematic lightcurve offsets. There is good evidence that the optical to near infrared spectra of most TNOs can be described by a single constant slope from the B through J bandpasses (Doressoundiram et al. 2007; Perna et al. 2010). In order to remove possible lightcurve effects, we therefore make the assumption that our targets likewise have a constant slope from F555W through F110W. We normalize the infrared measurements to the optical photometry by extrapolating the measured slope in the F555W, F675W and F814W filters to the effective wavelength of the F110W filter, 1.10 $\mu$m. We then shift the observed F110W and F160W photometry by an equal amount to force an agreement with the extrapolated value at 1.10 $\mu$m. The median absolute shift was 0.33 in normalized flux units (~1.5 magnitudes), and the values ranged from 2.53 to -0.54.

Particularly large offsets (>0.75 normalized flux units) were required for a small number of targets and may be indicative of objects with large amplitude lightcurves, unusual spectral features or phase functions. (19255) 1999 VK$_8$, 1999 OE$_4$, 2000 CE$_{105}$, 1998 KG$_{62}$ and 1998 WX$_{31}$ all met this criterion. Observations by Romanishin & Tegler (1999) found (19255) 1999 VK$_8$ to have a substantial lightcurve amplitude, 0.42 magnitudes although the period is not well known. None of the other objects have published lightcurve observations and all of these objects could be of interest for future studies.

## 5.    CONCLUSIONS

We have added 152 new color measurements to growing archive of TNO photometry. At optical wavelengths we have added 72 new measurements, ~30% of them of previously unobserved objects. At infrared wavelengths we have added 80 new measurements, ~80% of them previously unobserved. We have added an especially uniform and consistent sample to the database. We measure an average size population ~50 km smaller than previously studied, but do not find significant differences in their color properties. We find a >99.99% probability that the Cold and Hot/Resonant populations come from different distributions in the optical; we find a similar result in the infrared, but at a lower significance level. The offsets between the optical and infrared colors of (19255) 1999 $VK_8$, 1999 $OE_4$, 2000 $CE_{105}$, 1998 $KG_{62}$ and 1998 $WX_{31}$ are significantly different than expected, suggesting the presence of potentially large amplitude lightcurve effects, unusual spectral features, or phase functions.

### ACKNOWLEDGMENTS

This work is based on observations made with the NASA/ESA Hubble Space Telescope. These observations are associated with programs #9060, and 9386, some support for the analysis came from program #11178. Support for these programs was provided by NASA through a grant from the Space Telescope Science Institute, which is operated by the Association of Universities for Research in Astronomy, Inc., under NASA contract NAS 5-26555. We would also like to thank Francesca DeMeo and an anonymous reviewer for their helpful and constructive comments.

### REFERENCES

Anderson J. & King, I R. 1999. Astrometric and Photometric Corrections for the 34th Row Error in HST's WFPC2 Camera. The Publications of the Astronomical Society of the Pacific. 111, 1095-1098.

Baggett, S., and 22 colleagues, 2002. In: Mobasher, B. (Ed.), HST WFPC2 Data Handbook: Version 4.0. STScI, Baltimore.

Benecchi, S. D., Noll, K. S., Grundy, W. M., Buie, M. W., Stephens, D. C., Levison, H. F. 2009. The correlated colors of Transneptunian binaries. *Icarus* 200, 292-303.

Boehnhardt et al., 2002. ESO large program on physical studies of Transneptunian Objects and Centaurs: Visible photometry - First results. Astronomy and Astrophysics 395, 297-303.

Boehnhardt et al., 2003. Results from the Eso Large Program on Transneptunian Objects and Centaurs. Earth 92,145-156

Brucker et al., 2009. High albedos of low inclination Classical Kuiper belt objects. Icarus 201, 284-294.

Cook, J. C., Desch, S. J. and Rubin, M. 2011. The Black Sheep of Haumea's Collisional Family. 42nd Lunar and Planetary Science Conference 42, p. 2503.

Cox, A. N. 2000, Allen's Astrophysical Quantities (New York: Springer), p. 341.

de Jong, R. S., 2006. Correcting the NICMOS count-rate dependent non-linearity. Instrument Science Report NICMOS 2006-003, 3-14.

Delsanti et al., 2006. Near-Infrared Color Properties of Kuiper Belt Objects and Centaurs: Final Results from the ESO Large Program. The Astronomical Journal 131, 1851-1863.

Delsanti, A. C., Boehnhardt, H., Barrera, L., Meech, K. J., Sekiguchi, T, Hainaut, O. R. 2001. BVRI Photometry of 27 Kuiper Belt Objects with ESO/Very Large Telescope. Astronomy and Astrophysics 380, 347-358.

DeMeo, F. E., Fornasier, S., Barucci, M. A., Perna, D., Protopapa, S., Alvarez-Candal, A., Delsanti, A. C., Doressoundiram, A., Merlin, F., de Bergh, C. 2009. Visible and near-infrared colors of Transneptunian objects and Centaurs from the second ESO large program. Astronomy and Astrophysics 493, 283-290.

Dolphin. A. E. 2009. A Revised Characterization of the WFPC2 CTE Loss. Publications of the Astronomical Society of the Pacific 121, 655-667.

Dolphin. A. E. 2000. The Charge-Transfer Efficiency and Calibration of WFPC2. The Publications of the Astronomical Society of the Pacific 112, 1397-1410.

Doressoundiram, A., Boehnhardt, H., Tegler, S. C., and Trujillo, C. 2008. Color Properties and Trends of the Transneptunian Objects. In: M. A. Barucci, H. Boehnhardt, D. P. Cruikshank, and A. Morbidelli (Eds.), *The Solar System Beyond Neptune*, Univ. of Arizona Press, Tucson, pp. 91-104.

Doressoundiram et al., 2007. The Meudon Multicolor Survey (2MS) of Centaurs and Trans-Neptunian Objects: From Visible to Infrared Colors. The Astronomical Journal 134, 2186-2199.

Doressoundiram et al., 2005. The Meudon Multicolor Survey (2MS) of Centaurs and trans-neptunian objects: extended dataset and status on the correlations reported. Icarus 174, 90-104.

Doressoundiram, A, Barucci, M. A., Romon, J, Veillet, C. 2001. Multicolor Photometry of Trans-neptunian Objects. Icarus 154, 277-286.

Dressel, L., et al. 2007, "STIS Data Handbook", Version 5.0, (Baltimore: STScI).

Fernie. 1983. Relationships between the Johnson and Kron-Cousins VRI photometric systems. Astronomical Society of the Pacific. 95 pp. 782-785.

Gomes, R. S., Levison, H. F., Tsiganis, K., & Morbidelli, A. 2005. Origin of the cataclysmic Late Heavy Bombardment period of the terrestrial planets. Nature 435, 466-469.

Gulbis, A. A. S., Elliot, J. L., Kane, J. F. 2006. The color of the Kuiper belt Core. Icarus 183, 168-178.

Hainaut, O. R., Delsanti, A. C. 2002. Colors of Minor Bodies in the Outer Solar System. A statistical analysis. Astronomy and Astrophysics 389, 641-664.

Hawarden, T. G., Leggett, S. K., Letawsky, M. B., Ballantyne, D. R., Casali, M. M. 2001. JHK standard stars for large telescopes: the UKIRT Fundamental and Extended lists. Monthly Notices of the Royal Astronomical Society 325, 563-574.

Howell, S. B., Koehn, B., Bowell, E., Hoffman, M. 1996. Detection and Measurement of Poorly Sampled Point Sources Imaged With 2-D Array. Astronomical Journal 112, 1302-1311.

Hunt et al. 1998. Northern JHK Standard Stars for Array Detectors. The Astronomical Journal 115, 2594-2603.

Jewitt, D. and Luu, J. 2001. Colors and Spectra of Kuiper Belt Objects. The Astronomical Journal 122, 2099-2114.

Krist, J. & Hook R. 2004. The Tiny Tim User's Guide: Version 6.3 (Baltimore: STScI).

Levison, H. F., Morbidelli, A., Vanlaerhoven, C., Gomes, R. S., Tsiganis, K. 2008. Origin of the structure of the Kuiper belt during a dynamical instability in the orbits of Uranus and Neptune. Icarus 196, 258-273.

McMaster, Biretta, et al. 2008, WFPC2 Instrument Handbook, Version 10.0 (Baltimore: STScI).

Morbidelli, A., Levison, H. F., Tsiganis, K, Gomes, R. S. 2005. Chaotic capture of Jupiter's Trojan asteroids in the early Solar System. Nature 435, 462-465.

Noll, K. S., Grundy, W. M., Stephens, D. C., Levison, H. F., Kern, S. D. 2008. Evidence for two populations of classical transneptunian objects: The strong inclination dependence of classical binaries. Icarus 194, 758-768.

Noll, K., Stephens, D., Grundy, W., Cruikshank, D., Romanishin, W., Tegler, S. 2005. Infrared Photometry of Kuiper Belt Objects with NICMOS. American Astronomical Society 37, 746.

Pan, M., & Sari, R. 2005 Shaping the Kuiper belt size distribution by shattering large but strengthless bodies. Icarus, 173, 342-348.

Pavlovsky, C. 2006. ACS Data Handbook, Version 5.0 (Baltimore: Space Telescope Science Institute).

Peixinho, N., Lacerda, P., Jewitt, D. C. 2008.Color-Inclination Relation of the Classical Kuiper Belt Objects. The Astronomical Journal 136, 1837-1845.

Peixinho, N., Boehnhardt, H., Belskaya, I. N., Doressoundiram, A., Barucci, M. A., Delsanti, A. C. 2004. ESO large program on Centaurs and TNOs: visible colors-final results. Icarus 170, 153-166.

Peixinho, N., Doressoundiram, A., Delsanti, A. C., Boehnhardt, H., Barucci, M. A, and Belskaya, I. N. 2003. Reopening the TNOs color controversy: Centaurs bimodality and TNOs unimodality. Astronomy and Astrophysics 410, 29-32.

Perna et al. 2010. Colors and taxonomy of Centaurs and trans-Neptunian objects. Astronomy and Astrophysics 510, A53.

Persson, S. E., Murphy, D. C., Krzeminski, W., Roth, M., Rieke, M. J. 1998. A New System of Faint Near-Infrared Standard Stars. The Astronomical Journal 116, 2475-2488.

Ragozzine, D. & Brown, M. 2007. Candidate Members and Age Estimate of the Family of Kuiper Belt Object 2003 EL61. The Astronomical Journal 134, 2160-2167.

Romanishin, W. J., Tegler, S. C., Consolmagno, G. J. 2010. Colors of Inner Disk Classical Kuiper Belt Objects. The Astronomical Journal 140, 29-33.

Romanishin, W. J., & Tegler, S. C. 2005. Accurate absolute magnitudes for Kuiper belt objects and Centaurs. Icarus 179, 523-526.

Romanishin, W. J., & Tegler, S. C. 1999. Rotation rates of Kuiper-belt objects from their light curves. Nature 398, 129-132.

Santos-Sanz, P., Ortiz, J. L., Barrera, L, Boehnhardt, H. 2009. New BVRI photometry results on Kuiper Belt Objects from the ESO VLT. Astronomy and Astrophysics 494, 693-706.

Sheppard, S. S. 2010. The Colors of Extreme Outer Solar System Objects. The Astronomical Journal 139, 1394-1405.

Sheppard, S. S. 2007. Light Curves of Dwarf Plutonian Planets and other Large Kuiper Belt Objects: Their Rotations, Phase Functions, and Absolute Magnitudes. The Astronomical Journal 134, 787-798.

Sheppard, S. S., Jewitt, D. 2002. Time-resolved Photometry of Kuiper Belt Objects: Rotations, Shapes, and Phase Functions. The Astronomical Journal 124, 1757-1775.

Snodgrass, C., Carry, B., Dumas, C., Hainaut, O. R. 2010. Characterisation of candidate members of (136108) Haumea's family. Astronomy and Astrophysics 511, A72.

Stephens, D. C. & Noll, K. S. 2006. Detection of Six Trans-Neptunian Binaries with NICMOS: A High Fraction of Binaries in the Cold Classical Disk. The Astronomical Journal 131, 1142-1148.

Stephens, D. C. et al. 2003. HST Photometry of Transneptunian Objects. Earth, Moon and Planets 92, 251-260.

Tegler, S. C., Bauer, J. M., Romanishin, W., & Peixinho, N. 2008. Colors of Centaurs. In: M. A. Barucci, H. Boehnhardt, D. P. Cruikshank, and A. Morbidelli (Eds.), *The Solar System Beyond Neptune*, Univ. of Arizona Press, Tucson, pp.105-114.

Tegler, S. C., Romanishin, W. J., Consolmagno, G. J. 2003. Color Patterns in the Kuiper Belt: A Possible Primordial Origin. The Astrophysical Journal 599, 49-52.

Tegler, S. C., Romanishin, W. J. 2003. Resolution of the kuiper belt object color controversy: two distinct color populations. Icarus 161 pp. 181-191.

Tegler, S. C., Romanishin, W. J. 2000. Extremely red Kuiper-belt objects in near-circular orbits beyond 40 AU. Nature 407, 979-981.

Tegler, S. C., Romanishin, W. J. 1998. Two distinct populations of Kuiper-belt objects. Nature 392, 49-51.

Thatte, D. and Dahlen, T. et al. 2009, "NICMOS Data Handbook", version 8.0, (Baltimore, STScI)

Trujillo, C. and Brown, M. E. 2002. A Correlation between Inclination and Color in the Classical Kuiper Belt. The Astrophysical Journal 566, 125-128.

Tsiganis, K., Gomes, R. S., Morbidelli, A., Levison, H. F. 2005. Origin of the orbital architecture of the giant planets of the Solar System. Nature 435, 459-461.

Viana, A., Wiklind, T., et al. 2009, "NICMOS Instrument Handbook", Version 11.0, (Baltimore: STScI).

**FIGURES**

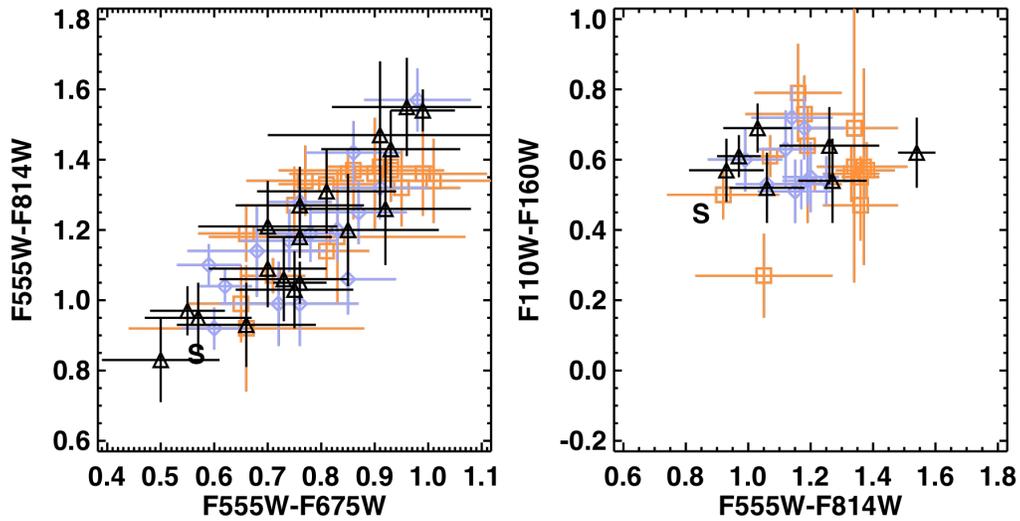

**Figure 1. HST Color-color plots.** The Sun is denoted as "S" in each plot. Objects are identified with respect to their general dynamical classifications with orange squares denoting Cold Classical objects, blue diamonds denoting Hot objects and black triangles denoting Resonant objects. Our optical colors span the full range of possible colors from 2000 JG$_{81}$ which is slightly bluer than the Sun to 1998 XY$_{95}$, 1993 SC and 1996 TQ$_{66}$, which are comparably very red. These data show no statistically significant differences between the colors of the three dynamical populations described in the text. The objects with overlapping optical and infrared colors are clustered around F110W-F160W of 0.58 with the exception of 2000 CE$_{105}$ which is significantly bluer.

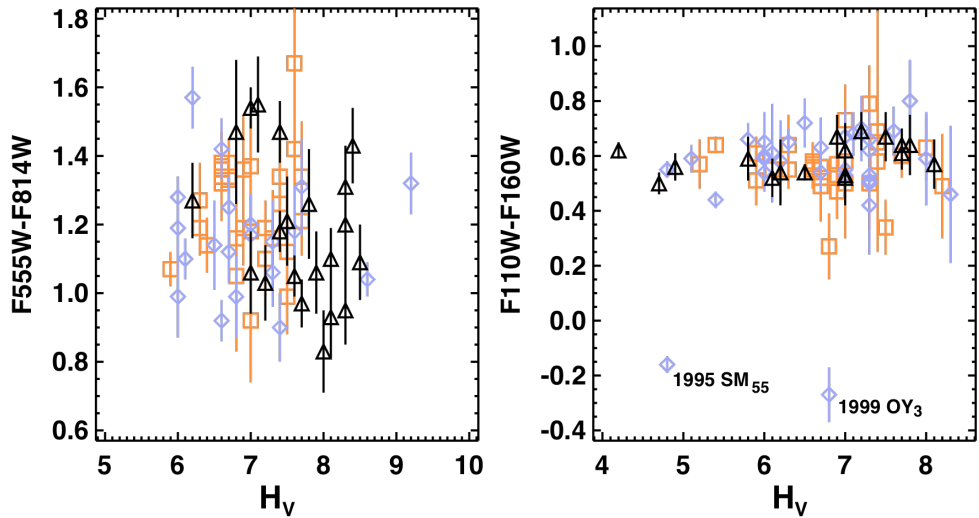

**Figure 2. Color vs. Hᵥ magnitude.** Objects are identified with respect to their general dynamical classifications with orange squares denoting Cold Classical objects, blue diamonds denoting Hot objects and black triangles denoting Resonant objects. These data show no statistically significant differences in the color distribution between dynamical class and $H_v$. The two outliers in the right panel, (86047) 1999 $OY_3$ and (24835) 1995 $SM_{55}$, have been identified as Haumea collision family members (Ragozzine & Brown, 2007). The colors of these objects are notably different than the colors of any other TNOs in these filters.

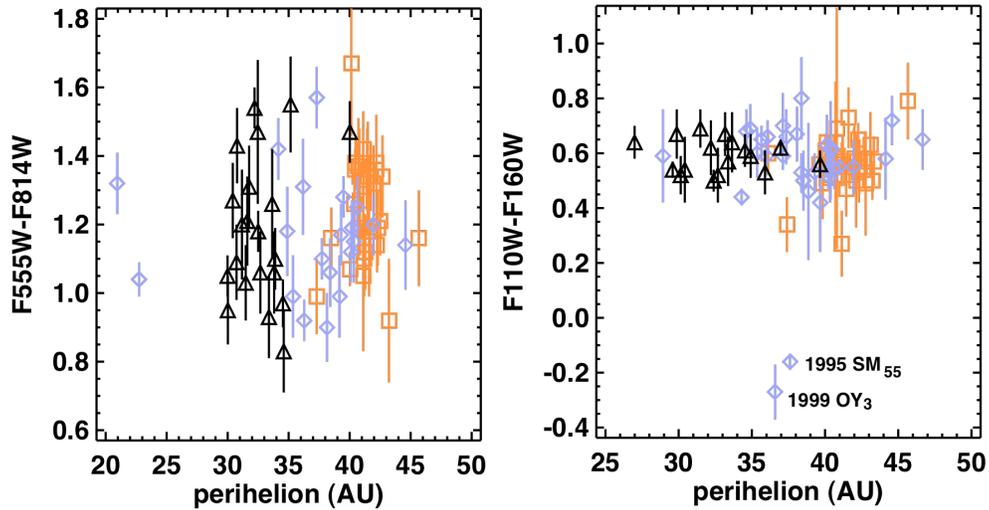

**Figure 3. Color vs. perihelion.** Objects are identified with respect to their general dynamical classifications with orange squares denoting Cold Classical objects, blue diamonds denoting Hot objects and black triangles denoting Resonant objects. These data show no statistically significant differences in the color distribution between dynamical class and perihelion. The two outliers in the right panel have been identified as Haumea family members (Ragozzine & Brown, 2007).

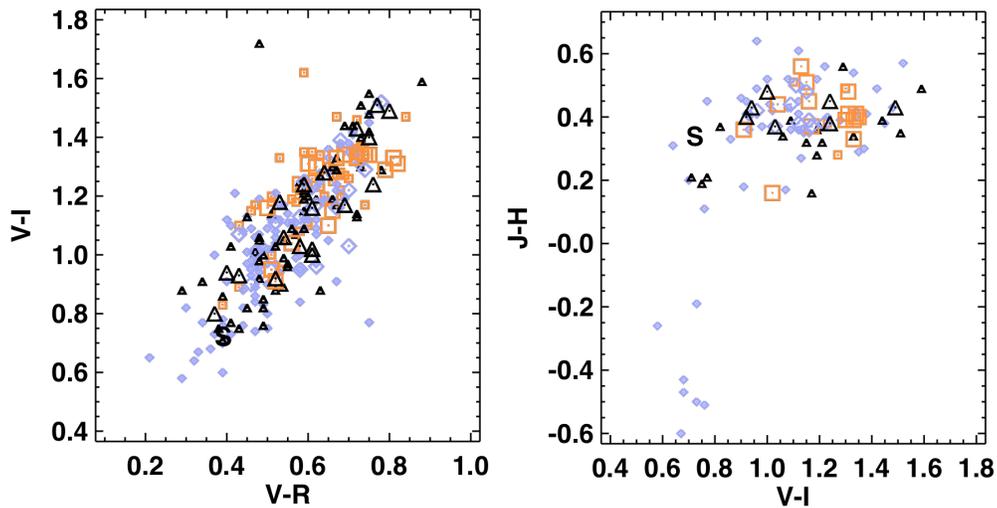

**Figure 4. HST and Literature colors.** Objects are identified with respect to their general dynamical classifications with orange squares denoting Cold Classical objects, blue diamonds denoting Hot objects and black triangles denoting Resonant objects, literature colors are identified with small symbols while HST colors are identified with large symbols. The Sun is denoted as "S" in each plot. The objects on the lower left part of the right plot belong to the Haumea collision family. Error bars are on the order of 0.05-0.15 magnitudes and have been left off the plot for clarity. Fundamentally, there are no differences between the quantitative results of these two datasets. Literature colors for this figure include the following datasets: Doressoundiram et al. (2001), Delsanti et al. 2001, Jewitt et al. 2001, Trujillo et al. (2002), Boehnhardt et al. 2002, Doressoundiram 2005 and 2007, Delsanti et al. (2006) Peixinho et al. 2004, Tegler, Romanishin, & Consolmagno database (http://www.physics.nau.edu/~tegler/research/survey.htm) inclusive of values from: Tegler & Romanishin 1998; Romanishin & Tegler 1999; Tegler & Romanishin 2000; Tegler & Romanishin 2003 and Tegler et al. 2003a), DeMeo et al. 2009, Santos-Sanz et al. 2009, Perna et al. (2010), Snodgrass et al. (2010), and Sheppard 2010. A copy of our compiled database, incorporating both this work and the cited references can be found in online supplemental material. The columns of the table include: (1) object number, (2) MPC designation, (3) MPC name, (4) DES classification, (5) MPC $H_V$, (6-23) three columns for each color: color, sigma on color, and the number of measurements combined, (24) combined list of references where the reference is defined by the first three letters of the first author's last name and the 2 year extension of the publication (this work is referenced as Ben11).

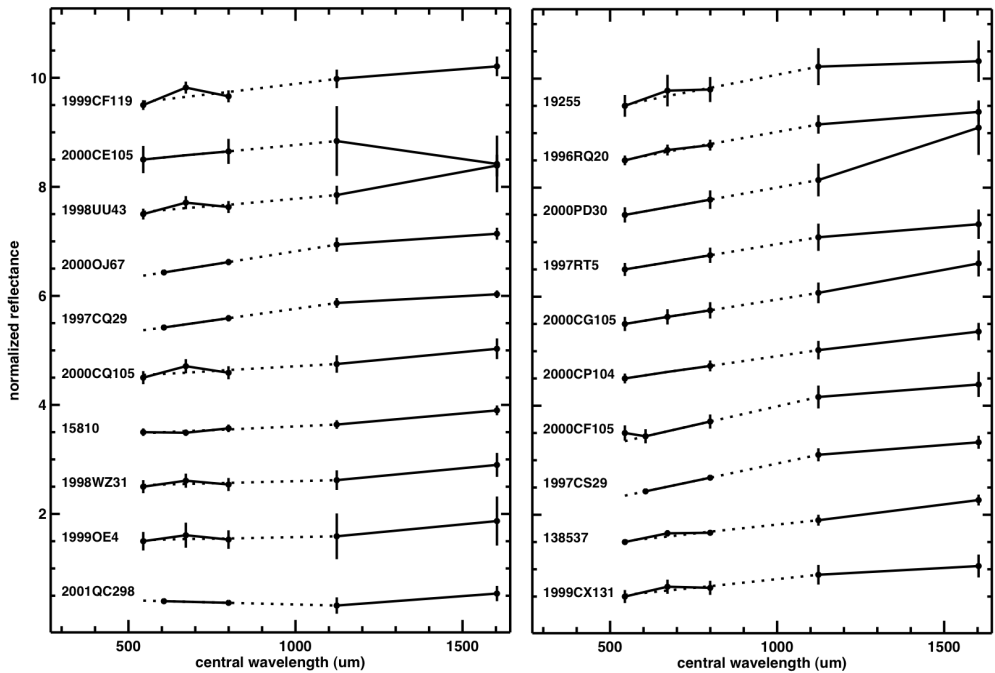

**Figure 5. Normalized reflectance measurements (blue).** The objects with measurements in 4 and 5 HST filters from this dataset are plotted vs. wavelength and normalized at F555W. Five additional objects with infrared colors in this dataset and optical colors from the Benecchi et al. (2009) dataset (F606W and F814W) are also plotted. An offset based on a linear fit to the optical

data for each object was applied to the F110W and F160W data to normalize all the measurements to a common baseline. Each spectrum is offset by 1 from the previous spectrum for clarity. We suggest that objects with large offsets, such as (19255) 1999 $VK_8$, 1999 $OE_4$, 2000 $CE_{105}$, 1998 $KG_{62}$ and 1998 $WX_{31}$ have substantial lightcurves or other interesting surface properties.

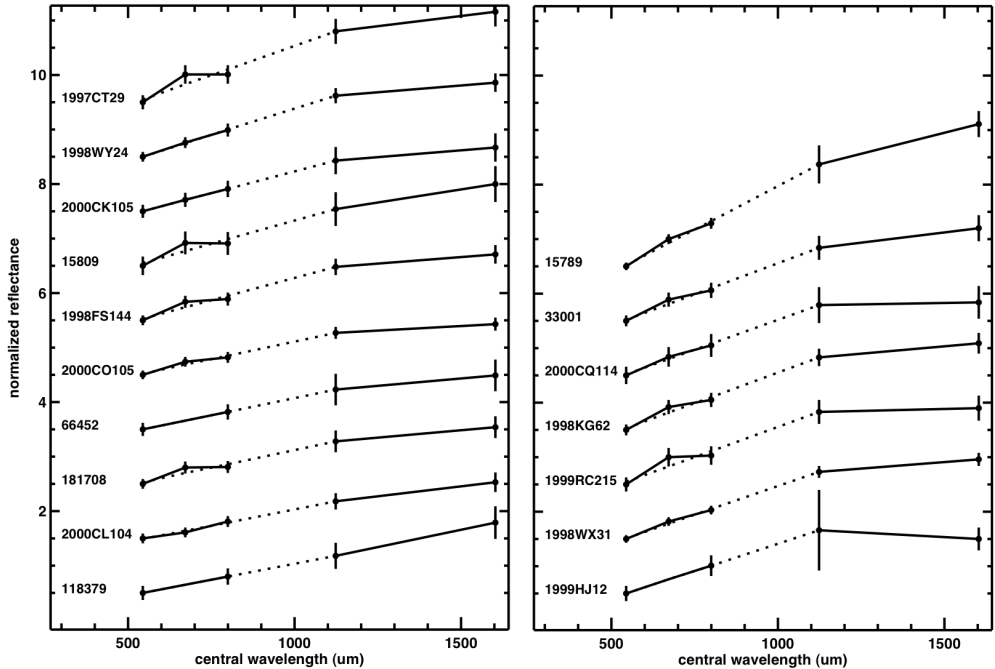

**Figure 6. Normalized reflectance measurements (red).** The same as for Figure 5, except for the objects with redder slopes.

**Table 1. Orbit and Observation Information**

| Number | Object | Julian Date[a] | Optical | | | Infrared | | | Class General[b] | DES Class[c] | GMV Class[d] | $a$[e] | $e$[e] | $i$[e] | $q$[e] |
|---|---|---|---|---|---|---|---|---|---|---|---|---|---|---|---|
| | | | $R_{\oplus}$ (AU) | Δ (AU) | α (°) | $R_{\oplus}$ (AU) | Δ (AU) | α (°) | | | | | | | |
| 181902 | 1999RD215 | 2452239.97668 | 39.15 | 38.51 | 1.110 | — | — | — | H | EX | DT[f] | 122.26 | 0.69 | 26.93 | 38.14 |
| 181874 | 1999HW11 | 2452317.55492 | 41.88 | 41.49 | 1.245 | — | — | — | H | EX | DT | 52.80 | 0.26 | 17.34 | 39.20 |
| 181871 | 1999CO153 | 2452371.61697 | 41.06 | 40.73 | 1.327 | — | — | — | R | 7:4 | CL[f] | 43.73 | 0.09 | 2.67 | 40.00 |
| 181708 | 1993FW | 2452325.56256 | 41.91 | 41.26 | 1.022 | — | — | — | H | CL | CL | 43.78 | 0.04 | 8.13 | 41.92 |
| — | — | 2452714.21490 | — | — | — | 41.89 | 41.00 | 0.612 | — | — | — | — | — | — | — |
| 168703 | 2000GP183 | 2452285.49311 | 37.28 | 37.44 | 1.488 | — | — | — | H | CL | CL | 39.81 | 0.09 | 11.89 | 36.26 |
| 168700 | 2000GE147 | 2452119.80907 | 30.31 | 30.62 | 1.821 | — | — | — | R | 3:2 | 3:2 | 39.45 | 0.22 | 5.54 | 30.72 |
| 148209 | 2000CR105 | 2452556.40166 | — | — | — | 53.93 | 54.47 | 0.886 | H | EX | DT | 221.83 | 0.80 | 21.36 | 44.22 |
| 138537 | 2000OK67 | 2452265.41881 | 40.77 | 41.24 | 1.208 | — | — | — | C | CL | CL | 46.56 | 0.14 | 5.00 | 40.04 |
| — | — | 2452553.09598 | — | — | — | 40.72 | 39.95 | 0.901 | — | — | — | — | — | — | — |
| 137294 | 1999RE215 | 2452163.97296 | 42.75 | 41.79 | 0.427 | — | — | — | C | CL | CL | 45.03 | 0.10 | 1.72 | 40.48 |
| 134860 | 2000OJ67 | 2452816.23323 | — | — | — | 42.65 | 42.01 | 1.070 | C | CL | CL | 42.90 | 0.01 | 1.32 | 42.35 |
| 131695 | 2001XS254 | 2452754.88785 | — | — | — | 35.28 | 35.58 | 1.554 | C | CL | 11:8[f] | 37.23 | 0.03 | 3.54 | 36.11 |
| 130391 | 2000JG81 | 2452122.89171 | 34.29 | 34.19 | 1.690 | — | — | — | R | 2:1 | 2:1 | 47.80 | 0.28 | 23.03 | 34.58 |
| 119070 | 2001KP77 | 2452549.81444 | — | — | — | 36.04 | 36.65 | 1.249 | R | 7:4 | 7:4 | 43.73 | 0.18 | 2.77 | 35.91 |
| 118379 | 1999HC12 | 2452318.49265 | 39.78 | 39.61 | 1.405 | — | — | — | H | EX | CL | 45.35 | 0.23 | 14.45 | 34.89 |
| — | — | 2452773.98442 | — | — | — | 40.04 | 39.04 | 0.216 | — | — | — | — | — | — | — |
| 118228 | 1996TQ66 | 2452310.23270 | 34.61 | 34.80 | 1.597 | — | — | — | R | 3:2 | 3:2 | 39.46 | 0.11 | 14.81 | 35.16 |
| 91133 | 1998HK151 | 2452157.07435 | 30.44 | 30.70 | 1.824 | — | — | — | R | 3:2 | 3:2 | 39.45 | 0.24 | 4.47 | 30.00 |
| 87269 | 2000OO67 | 2452444.34174 | 21.13 | 20.83 | 2.650 | — | — | — | H | CN | SC | 510.75 | 0.96 | 19.26 | 21.00 |
| 86177 | 1999RY214 | 2452758.47214 | — | — | — | 36.64 | 37.06 | 1.429 | H | EX | CL | 45.31 | 0.24 | 23.34 | 34.60 |
| 86047[g] | 1999OY3 | 2452556.62897 | — | — | — | 39.34 | 38.98 | 1.366 | H | EX | CL | 43.92 | 0.17 | 25.82 | 36.59 |
| 85633 | 1998KR65 | 2452155.13894 | 44.31 | 43.47 | 0.725 | — | — | — | C | CL | CL | 43.56 | 0.04 | 0.94 | 41.74 |
| 85627 | 1998HP151 | 2452170.57043 | 42.60 | 43.12 | 1.145 | — | — | — | C | CL | CL | 44.04 | 0.08 | 1.01 | 40.40 |
| 82075 | 2000YW134 | 2452572.52417 | — | — | — | 42.96 | 42.94 | 1.327 | R | 8:3 | 8:3 | 57.90 | 0.32 | 16.68 | 39.69 |
| 80806 | 2000CM105 | 2452552.70999 | — | — | — | 41.84 | 42.45 | 1.085 | C | CL | CL | 42.20 | 0.06 | 2.13 | 39.83 |
| 79360 | 1997CS29 | 2452570.48637 | — | — | — | 43.55 | 43.60 | 1.307 | C | CL | CL | 43.91 | 0.01 | 3.84 | 43.36 |
| 69988 | 1998WA31 | 2452156.29171 | 39.65 | 39.16 | 1.283 | — | — | — | R | 5:2 | 5:2 | 55.46 | 0.43 | 9.44 | 31.64 |
| 69987 | 1998WA25 | 2452164.11557 | 42.05 | 41.62 | 1.250 | — | — | — | C | CL | — | 42.56 | 0.03 | 2.13 | 41.21 |
| 66652 | BORASISI | 2452752.64328 | — | — | — | 41.07 | 41.56 | 1.217 | C | CL | — | 43.84 | 0.08 | 1.57 | 40.31 |
| 66452 | 1999OF4 | 2452368.52044 | 45.12 | 45.71 | 1.023 | — | — | — | C | CL | CL | 45.01 | 0.06 | 1.80 | 42.33 |

| Number | Object | Julian Date[a] | Optical | | | Infrared | | | Class General[b] | DES Class[c] | GMV Class[d] | $a$[e] | $e$[e] | $i$[e] | $q$[e] |
|---|---|---|---|---|---|---|---|---|---|---|---|---|---|---|---|
| | | | $R_\oplus$ (AU) | $\Delta$ (AU) | $\alpha$ (°) | $R_\oplus$ (AU) | $\Delta$ (AU) | $\alpha$ (°) | | | | | | | |
| — | — | 2452553.90646 | — | — | — | 45.15 | 44.56 | 1.027 | — | — | — | — | — | — | — |
| 60621 | 2000FE8 | 2452621.38258 | — | — | — | 35.95 | 36.30 | 1.457 | R | 5:2 | 5:2 | 55.46 | 0.40 | 6.73 | 33.18 |
| 60620 | 2000FD8 | 2452239.54242 | 39.92 | 40.47 | 1.168 | — | — | — | H | EX | 7:4 | 43.72 | 0.22 | 19.40 | 34.14 |
| 60608 | 2000EE173 | 2452322.00145 | 23.12 | 22.13 | 0.154 | — | — | — | H | CN | SC | 49.55 | 0.54 | 8.28 | 22.75 |
| 60454 | 2000CH105 | 2452290.56742 | 43.87 | 43.21 | 0.969 | — | — | — | C | CL | CL | 44.38 | 0.08 | 2.81 | 40.83 |
| 58534 | LOGOS | 2452763.70253 | — | — | — | 41.69 | 41.19 | 1.208 | C | CL | CL | 45.28 | 0.13 | 2.02 | 39.62 |
| 54520 | 2000PJ30 | 2452563.40073 | — | — | — | 41.01 | 40.74 | 1.342 | H | CN | SC | 130.31 | 0.78 | 8.22 | 28.92 |
| 48639 | 1995TL8 | 2452587.74656 | — | — | — | 42.72 | 41.76 | 0.333 | C | CL | CL | 52.58 | 0.24 | 1.88 | 40.11 |
| 45802 | 2000PV29 | 2452556.76292 | — | — | — | 43.31 | 42.94 | 1.232 | C | CL | CL | 43.50 | 0.01 | 2.19 | 43.11 |
| — | — | 2452570.64565 | — | — | — | 43.31 | 43.17 | 1.305 | — | — | — | — | — | — | — |
| 42301 | 2001UR163 | 2452664.14694 | — | — | — | 49.06 | 49.28 | 1.118 | R | 9:4 | 9:4 | 51.70 | 0.29 | 2.42 | 36.97 |
| 38084 | 1999HB12 | 2452273.47506 | 34.70 | 34.82 | 1.609 | — | — | — | R | 5:2 | 5:2 | 56.46 | 0.41 | 12.38 | 32.52 |
| 33340 | 1998VG44 | 2452586.88382 | — | — | — | 30.04 | 29.10 | 0.636 | R | 3:2 | 3:2 | 39.46 | 0.25 | 2.03 | 29.59 |
| 33001 | 1997CU29 | 2452216.00492 | 44.68 | 44.48 | 1.250 | — | — | — | C | CL | CL | 43.50 | 0.03 | 2.50 | 42.20 |
| — | — | 2452553.53396 | — | — | — | 44.66 | 44.96 | 1.221 | — | — | CL | — | — | — | — |
| 26375 | 1999DE9 | 2452793.98830 | — | — | — | 34.48 | 34.52 | 1.683 | R | 5:2 | 5:2 | 55.46 | 0.42 | 9.17 | 32.39 |
| 24978 | 1998HJ151 | 2452324.63223 | 42.14 | 42.07 | 1.341 | — | — | — | C | CL | CL | 43.37 | 0.05 | 1.45 | 41.28 |
| 24835# | 1995SM55 | 2452549.63149 | — | — | — | 39.22 | 38.30 | 0.606 | H | EX | CL | 41.85 | 0.10 | 26.79 | 37.63 |
| 20161 | 1996TR66 | 2452599.62167 | — | — | — | 37.85 | 37.18 | 1.099 | R | 2:1 | 2:1 | 47.79 | 0.38 | 15.48 | 29.88 |
| 20108 | 1995QZ9 | 2452160.23432 | 35.28 | 34.47 | 0.966 | — | — | — | R | 3:2 | 3:2 | 39.46 | 0.14 | 19.64 | 33.82 |
| 19521 | CHAOS | 2452575.35183 | — | — | — | 42.16 | 41.28 | 0.641 | R | CL | CL | 45.78 | 0.11 | 11.00 | 40.96 |
| 19255 | 1994VK8 | 2452123.53894 | 43.40 | 43.93 | 1.143 | — | — | — | C | CL | CL | 42.84 | 0.03 | 1.11 | 41.61 |
| — | — | 2452525.55169 | — | — | — | 43.38 | 43.32 | 1.330 | — | — | — | — | — | — | — |
| 16684 | 1994JQ1 | 2452267.67506 | 42.85 | 43.45 | 1.038 | — | — | — | C | CL | CL | 44.13 | 0.04 | 3.39 | 42.21 |
| 15883 | 1997CR29 | 2452566.35492 | — | — | — | 42.59 | 43.05 | 1.180 | H | EX | CL | 47.05 | 0.21 | 17.85 | 37.11 |
| 15874 | 1996TL66 | 2452581.14787 | — | — | — | 35.03 | 34.05 | 0.164 | H | SN | SC | 83.65 | 0.59 | 23.49 | 34.28 |
| 15836 | 1995DA2 | 2452290.09196 | 34.20 | 33.32 | 0.728 | — | — | — | R | 4:3 | 4:3 | 36.48 | 0.07 | 4.83 | 33.88 |
| 15810 | 1994JR1 | 2452118.61741 | 34.83 | 34.16 | 1.269 | — | — | — | R | 3:2 | 3:2 | 39.46 | 0.13 | 3.42 | 34.54 |
| — | — | 2452550.75495 | — | — | — | 34.85 | 35.22 | 1.524 | — | — | — | — | — | — | — |
| 15809 | 1994JS | 2452188.62921 | 34.81 | 35.36 | 1.353 | — | — | — | R | 5:3 | 5:3 | 42.33 | 0.21 | 13.51 | 33.63 |
| — | — | 2452556.54682 | — | — | — | 34.66 | 35.22 | 1.346 | — | — | — | — | — | — | — |
| 15789 | 1993SC | 2452122.04657 | 35.09 | 34.67 | 1.521 | — | — | — | R | 3:2 | 3:2 | 39.46 | 0.18 | 5.89 | 32.19 |
| — | — | 2452551.89973 | — | — | — | 35.29 | 34.29 | 0.144 | — | — | — | — | — | — | — |

                                                                                   

| Number | Object | Julian Date[a] | Optical | | | Infrared | | | Class General[b] | DES Class[c] | GMV Class[d] | $a^e$ | $e^e$ | $i^e$ | $q^e$ |
|---|---|---|---|---|---|---|---|---|---|---|---|---|---|---|---|
| | | | $R_\oplus$ (AU) | $\Delta$ (AU) | $\alpha$ (°) | $R_\oplus$ (AU) | $\Delta$ (AU) | $\alpha$ (°) | | | | | | | |
| 15788 | 1993SB | 2452578.07257 | — | — | — | 29.90 | 28.96 | 0.628 | R | 3:2 | 3:2 | 39.45 | 0.32 | 3.13 | 26.99 |
| 15760 | 1992QB1 | 2452122.44751 | 40.92 | 40.48 | 1.288 | — | — | — | C | CL | CL | 43.98 | 0.07 | 2.67 | 40.85 |
| — | 1995DB2 | 2452274.26834 | 40.22 | 39.47 | 0.917 | — | — | — | C | CL | CL | 46.51 | 0.14 | 2.86 | 40.12 |
| — | 1995HM5 | 2452267.47784 | 31.54 | 31.86 | 1.680 | — | — | — | R | 3:2 | 3:2 | 39.45 | 0.24 | 6.65 | 30.02 |
| — | 1996KV1 | 2452550.95382 | — | — | — | 40.56 | 40.87 | 1.339 | H | CL | CL | 45.25 | 0.11 | 6.19 | 40.42 |
| — | 1996RQ20 | 2452229.81950 | 39.48 | 38.69 | 0.867 | — | — | — | H | EX | CL | 43.90 | 0.11 | 31.71 | 39.27 |
| — | — | 2452551.56603 | — | — | — | 39.50 | 38.51 | 0.248 | — | — | — | — | — | — | — |
| — | 1996RR20 | 2452217.05492 | 43.91 | 43.32 | 1.042 | — | — | — | R | 3:2 | 3:2 | 39.46 | 0.18 | 4.96 | 32.48 |
| — | 1996TK66 | 2452246.05631 | 42.91 | 42.41 | 1.145 | — | — | — | C | CL | CL | 42.71 | 0.01 | 2.02 | 42.22 |
| — | 1997CT29 | 2452227.05006 | 44.80 | 44.78 | 1.265 | — | — | — | C | CL | CL | 43.79 | 0.03 | 0.98 | 42.70 |
| — | — | 2452679.78285 | — | — | — | 44.78 | 43.80 | 0.025 | — | — | — | — | — | — | — |
| — | 1997CV29 | 2452564.62005 | — | — | — | 40.24 | 40.77 | 1.192 | H | CL | CL[f] | 42.33 | 0.09 | 6.06 | 38.52 |
| — | 1997RT5 | 2452280.04890 | 42.35 | 42.78 | 1.186 | — | — | — | H | CL | CL | 41.37 | 0.03 | 12.69 | 40.34 |
| — | — | 2452551.83014 | — | — | — | 42.35 | 41.41 | 0.490 | — | — | — | — | — | — | — |
| — | 1998FS144 | 2452313.06256 | 41.91 | 41.10 | 0.786 | — | — | — | H | CL | CL | 41.82 | 0.03 | 11.60 | 40.60 |
| — | — | 2452809.59196 | — | — | — | 41.94 | 42.00 | 1.384 | — | — | — | — | — | — | — |
| — | 1998KG62 | 2452158.08060 | 45.20 | 44.39 | 0.760 | — | — | — | C | CL | CL | 43.46 | 0.06 | 2.12 | 40.81 |
| — | — | 2452570.71258 | — | — | — | 45.16 | 45.02 | 1.251 | — | — | — | — | — | — | — |
| — | 1998KS65 | 2452156.87668 | 42.31 | 41.48 | 0.786 | — | — | — | C | CL | CL | 43.83 | 0.03 | 0.91 | 42.52 |
| — | 1998KY61 | 2452761.47274 | — | — | — | 46.39 | 46.36 | 1.245 | C | CL | CL | 44.39 | 0.05 | 1.14 | 42.13 |
| — | 1998UR43 | 2452315.57645 | 31.49 | 31.64 | 1.772 | — | — | — | R | 3:2 | 3:2 | 39.45 | 0.21 | 8.08 | 31.19 |
| — | 1998UU43 | 2452321.31603 | 37.33 | 37.59 | 1.458 | — | — | — | R | 4:3 | 4:3 | 36.48 | 0.14 | 11.19 | 31.48 |
| — | — | 2452630.72588 | — | — | — | 37.22 | 36.55 | 1.118 | R | 4:3 | 4:3 | — | — | — | — |
| — | 1998WS31 | 2452146.74241 | 31.45 | 31.16 | 1.771 | — | — | — | R | 3:2 | 3:2 | 39.46 | 0.20 | 6.87 | 31.77 |
| — | 1998WV24 | 2452139.58130 | 38.30 | 38.18 | 1.507 | — | — | — | C | CL | CL | 39.02 | 0.04 | 1.95 | 37.30 |
| — | 1998WX24 | 2452320.85145 | 45.10 | 45.10 | 1.255 | — | — | — | C | CL | CL | 43.37 | 0.03 | 1.26 | 41.94 |
| — | 1998WX31 | 2452241.53617 | 40.59 | 39.61 | 0.166 | — | — | — | C | CL | CL | 45.49 | 0.11 | 2.38 | 40.45 |
| — | — | 2452527.62253 | — | — | — | 40.59 | 40.53 | 1.421 | — | — | — | — | — | — | — |
| — | 1998WY24 | 2452240.46603 | 41.77 | 40.82 | 0.356 | — | — | — | C | CL | CL | 43.25 | 0.05 | 1.02 | 41.13 |
| — | — | 2452606.90890 | — | — | — | 41.75 | 40.80 | 0.360 | — | — | — | — | — | — | — |
| — | 1998WZ31 | 2452148.47227 | 33.14 | 33.12 | 1.747 | — | — | — | R | 3:2 | 3:2 | 39.46 | 0.15 | 13.66 | 33.39 |
| — | — | 2452561.91957 | — | — | — | 33.21 | 32.49 | 1.213 | — | — | — | — | — | — | — |
| — | 1998XY95 | 2452150.28130 | 42.20 | 42.53 | 1.291 | — | — | — | H | EX | DT[f] | 64.44 | 0.42 | 6.24 | 37.33 |

| Number | Object | Julian Date[a] | Optical | | | Infrared | | | Class General[b] | DES Class[c] | GMV Class[d] | $a^e$ | $e^e$ | $i^e$ | $q^e$ |
|---|---|---|---|---|---|---|---|---|---|---|---|---|---|---|---|
| | | | $R_\oplus$ (AU) | $\Delta$ (AU) | $\alpha$ (°) | $R_\oplus$ (AU) | $\Delta$ (AU) | $\alpha$ (°) | | | | | | | |
| — | 1999CD158 | 2452687.71821 | — | — | — | 48.11 | 47.22 | 0.514 | H | EX | 7:4[f] | 43.73 | 0.15 | 23.78 | 37.12 |
| — | 1999CF119 | 2452258.43270 | 39.25 | 38.54 | 1.016 | — | — | — | H | EX | DT | 88.87 | 0.57 | 20.72 | 38.35 |
| — | — | 2452676.04559 | — | — | — | 39.11 | 38.13 | 0.197 | — | — | — | — | — | — | — |
| — | 1999CH119 | 2452571.62499 | — | — | — | 46.77 | 46.90 | 1.207 | H | EX | CL | 43.36 | 0.09 | 18.41 | 39.66 |
| — | 1999CJ119 | 2452671.91005 | — | — | — | 42.33 | 41.35 | 0.019 | C | CL | CL | 45.40 | 0.07 | 4.63 | 42.28 |
| — | 1999CL119 | 2452570.41716 | — | — | — | 46.71 | 46.91 | 1.193 | H | EX | CL | 46.94 | 0.01 | 21.81 | 46.64 |
| — | 1999CQ133 | 2452755.35808 | — | — | — | 43.78 | 43.77 | 1.317 | H | CL | CL | 41.30 | 0.08 | 11.38 | 38.08 |
| — | 1999CX131 | 2452272.47159 | 42.50 | 41.70 | 0.793 | — | — | — | R | 5:3 | 5:3 | 42.33 | 0.23 | 9.12 | 32.67 |
| — | — | 2452564.55091 | — | — | — | 42.32 | 42.68 | 1.252 | — | — | — | — | — | — | — |
| — | 1999HJ12 | 2452158.00584 | 44.13 | 44.64 | 1.117 | — | — | — | C | CL | CL | 43.09 | 0.05 | 3.54 | 40.80 |
| — | — | 2452527.57228 | — | — | — | 44.09 | 44.65 | 1.081 | — | — | — | — | — | — | — |
| — | 1999OD4 | 2452562.77043 | — | — | — | 44.30 | 44.03 | 1.247 | H | EX | CL | 41.47 | 0.10 | 15.77 | 37.37 |
| — | 1999OE4 | 2452369.58340 | 43.45 | 44.03 | 1.073 | — | — | — | C | CL | CL | 45.41 | 0.05 | 1.39 | 43.24 |
| — | — | 2452497.61306 | — | — | — | 43.45 | 42.43 | 0.043 | — | — | — | — | — | — | — |
| — | 1999OH4 | 2452762.47281 | — | — | — | 39.01 | 38.96 | 1.479 | H | EX | CL | 40.52 | 0.04 | 26.69 | 38.84 |
| — | 1999OJ4 | 2452552.09267 | — | — | — | 38.18 | 37.57 | 1.194 | C | CL | CL | 38.08 | 0.02 | 2.58 | 37.41 |
| — | 1999RC215 | 2452269.50075 | 43.29 | 43.20 | 1.298 | — | — | — | C | CL | | 44.09 | 0.06 | 2.42 | 41.44 |
| — | — | 2452589.81104 | — | — | — | 43.25 | 42.47 | 0.815 | — | — | — | — | — | — | — |
| — | 1999RX214 | 2452267.96418 | 45.94 | 45.86 | 1.223 | — | — | — | C | CL | — | 46.19 | 0.17 | 5.25 | 38.47 |
| — | 1999TR11 | 2452167.32644 | 30.00 | 29.62 | 1.789 | — | — | — | R | 3:2 | | 39.45 | 0.22 | 17.05 | 30.79 |
| — | 1999XY143 | 2452236.04797 | 43.58 | 42.65 | 0.441 | — | — | — | H | CL | CL | 43.06 | 0.08 | 9.11 | 39.46 |
| — | 2000AF255 | 2452587.09483 | — | — | — | 52.81 | 52.28 | 0.913 | H | EX | DT | 48.63 | 0.26 | 30.42 | 36.09 |
| — | 2000CE105 | 2452160.66348 | 41.39 | 41.97 | 1.132 | — | — | — | C | CL | CL | 43.94 | 0.06 | 1.23 | 41.14 |
| — | — | 2452705.32567 | — | — | — | 41.40 | 40.81 | 1.120 | — | — | — | — | — | — | — |
| — | 2000CF105 | 2452694.65606 | — | — | — | 42.28 | 41.42 | 0.688 | C | CL | CL | 43.91 | 0.04 | 1.36 | 42.00 |
| — | 2000CG105 | 2452288.56742 | 46.51 | 45.60 | 0.463 | — | — | — | H | EX | CL | 46.38 | 0.04 | 29.38 | 44.57 |
| — | — | 2452757.29400 | — | — | — | 46.55 | 46.39 | 1.225 | — | — | — | — | — | — | — |
| — | 2000CK105 | 2452277.41047 | 48.50 | 47.66 | 0.614 | — | — | — | R | 3:2 | 3:2 | 39.45 | 0.23 | 9.68 | 30.43 |
| — | — | 2452570.55529 | — | — | — | 48.52 | 48.78 | 1.130 | — | — | — | — | — | — | — |
| — | 2000CL104 | 2452290.50075 | 42.71 | 41.93 | 0.819 | — | — | — | C | CL | CL | 44.48 | 0.08 | 1.36 | 40.74 |
| — | — | 2452793.91938 | — | — | — | 42.80 | 42.94 | 1.342 | — | — | — | — | — | — | — |
| — | 2000CO105 | 2452294.36742 | 49.53 | 48.55 | 0.062 | — | — | — | H | EX | CL | 47.12 | 0.15 | 20.76 | 40.19 |
| — | — | 2452695.65627 | — | — | — | 49.66 | 48.84 | 0.633 | — | — | — | — | — | — | — |

| Number | Object | Julian Date[a] | Optical | | | Infrared | | | Class General[b] | DES Class[c] | GMV Class[d] | $a^e$ | $e^e$ | $i^e$ | $q^e$ |
|---|---|---|---|---|---|---|---|---|---|---|---|---|---|---|---|
| | | | $R_\odot$ (AU) | $\Delta$ (AU) | $\alpha$ (°) | $R_\odot$ (AU) | $\Delta$ (AU) | $\alpha$ (°) | | | | | | | |
| — | 2000CP104 | 2452313.52598 | 46.65 | 45.69 | 0.269 | — | — | — | H | CL | CL | 44.33 | 0.10 | 8.12 | 40.10 |
| — | — | 2452810.85468 | — | — | — | 46.55 | 47.00 | 1.115 | — | — | — | — | — | — | — |
| — | 2000CQ105 | 2452241.46881 | 50.28 | 49.98 | 1.074 | — | — | — | H | EX | DT | 57.41 | 0.38 | 19.21 | 35.35 |
| — | — | 2452551.70847 | — | — | — | 49.98 | 50.60 | 0.900 | — | — | — | — | — | — | — |
| — | 2000CQ114 | 2452280.62784 | 45.27 | 44.68 | 0.998 | — | — | — | C | CL | CL | 46.15 | 0.12 | 2.20 | 40.69 |
| — | — | 2452796.51789 | — | — | — | 45.42 | 45.53 | 1.271 | — | — | — | — | — | — | — |
| — | 2000FS53 | 2452118.13987 | 42.11 | 42.23 | 1.370 | — | — | — | C | CL | CL | 43.12 | 0.04 | 1.93 | 41.47 |
| — | 2000GV146 | 2452119.87668 | 41.76 | 41.94 | 1.368 | — | — | — | C | CL | CL | 44.08 | 0.07 | 2.04 | 41.12 |
| — | 2000KK4 | 2452118.07921 | 44.27 | 43.84 | 1.192 | — | — | — | H | EX | CL | 41.33 | 0.09 | 18.21 | 37.75 |
| — | 2000KL4 | 2452300.33640 | 40.18 | 40.63 | 1.242 | — | — | — | H | EX | CL | 38.65 | 0.06 | 20.42 | 36.17 |
| — | 2000OU69 | 2452557.29624 | — | — | — | 41.11 | 40.85 | 1.349 | C | CL | CL | 43.24 | 0.04 | 2.47 | 41.35 |
| — | 2000PD30 | 2452368.58086 | 45.72 | 46.29 | 1.020 | — | — | — | C | CL | CL | 46.60 | 0.02 | 5.75 | 45.68 |
| — | — | 2452766.54149 | — | — | — | 45.73 | 45.81 | 1.259 | — | — | — | — | — | — | — |
| — | 2000PE30 | 2452556.69611 | — | — | — | 37.27 | 36.91 | 1.439 | H | EX | DT | 54.52 | 0.35 | 16.63 | 35.67 |
| — | 2000PH30 | 2452562.83730 | — | — | — | 46.46 | 46.20 | 1.192 | H | EX | DT[f] | 76.94 | 0.50 | 6.77 | 38.36 |
| — | 2001KD77 | 2452550.88490 | — | — | — | 35.34 | 35.83 | 1.409 | R | 3:2 | 3:2 | 39.46 | 0.12 | 2.41 | 34.91 |
| — | 2001KY76 | 2452550.81563 | — | — | — | 38.73 | 39.18 | 1.309 | R | 3:2 | 3:2 | 39.45 | 0.24 | 1.64 | 30.15 |
| — | 2001OG109 | 2452764.47274 | — | — | — | 43.04 | 42.96 | 1.339 | C | CL | CL | 43.56 | 0.02 | 2.10 | 42.77 |
| — | 2001OK108 | 2452758.40157 | — | — | — | 42.26 | 42.25 | 1.365 | C | CL | CL | 42.99 | 0.02 | 1.41 | 41.99 |
| — | 2001QC298 | 2452552.16159 | — | — | — | 40.57 | 39.73 | 0.781 | H | EX | CL | 46.32 | 0.13 | 31.54 | 40.38 |
| — | 2001QX322 | 2452664.74218 | — | — | — | 39.11 | 39.49 | 1.326 | H | EX | — | 58.14 | 0.38 | 30.22 | 36.01 |
| — | 2001XU254 | 2452755.55479 | — | — | — | 42.57 | 42.89 | 1.282 | C | CL | CL | 43.49 | 0.07 | 5.00 | 40.36 |

[a] midtime of observation.

[b] C=Classical with $i\leq6°$, H=Classical with $i>6°$ + Scattered/Scattering + Detached + Centaurs, R=Resonant.

[c] CL=Classical, CN=Centaur, EX=Scattered extended, SN=Scattered near, n:m=resonance. Classifications from Elliot et al. 2005 modified to more closely align with, but not matching, Gladman et al. 2008 (Marc Buie private communication).

[d] GMV classifications: CL=Classical, SC=Scattered, DT=Detached, n:m=resonance.

[e] Mean values based on a 10 My integration of the orbit of the object.

[f] Tentative classification

[g] Haumea family object.

## Table 2. Photometry (HST Filters)

| Number | Object | F555W | F675W | F814W | F110W | F160W | F555W-F675W | F555W_F814W | F675W-F814W | F110W-F160W | $S^b$ |
|---|---|---|---|---|---|---|---|---|---|---|---|
| | Sun[a] | -26.73±0.01 | -27.28±0.01 | -27.55±0.01 | -27.73±0.01 | -28.15±0.01 | 0.55±0.01 | 0.82±0.01 | 0.27±0.01 | 0.42±0.01 | — |
| 181902 | 1999RD215 | 23.63±0.07 | — | 22.73±0.07 | — | — | — | 0.90±0.10 | — | — | 0.11±0.01 |
| 181874 | 1999HW11 | 23.57±0.09 | 22.85±0.07 | 22.59±0.08 | — | — | 0.72±0.11 | 0.99±0.12 | 0.26±0.11 | — | 1.88±0.22 |
| 181871 | 1999CO153 | 23.94±0.08 | — | 22.47±0.05 | — | — | — | 1.47±0.09 | — | — | 16.1±1.7 |
| 181708 | 1993FW | 23.41±0.07 | 22.58±0.06 | 22.22±0.06 | 21.63±0.08 | 21.08±0.06 | 0.83±0.09 | 1.20±0.09 | 0.36±0.08 | 0.55±0.10 | 7.21±0.66 |
| 168703 | 2000GP183 | 22.03±0.04 | 21.43±0.04 | 21.11±0.04 | — | — | 0.60±0.06 | 0.92±0.06 | 0.32±0.06 | — | 0.54±0.03 |
| 168700 | 2000GE147 | 23.41±0.08 | 22.71±0.07 | 22.32±0.07 | — | — | 0.70±0.11 | 1.09±0.11 | 0.39±0.10 | — | 4.54±0.47 |
| 148209 | 2000CR105 | — | — | — | 22.64±0.12 | 22.06±0.09 | — | — | — | 0.58±0.15 | — |
| 138537 | 2000OK67 | 22.74±0.04 | 22.03±0.04 | 21.67±0.03 | 21.08±0.04 | 20.46±0.03 | 0.71±0.06 | 1.07±0.05 | 0.36±0.05 | 0.61±0.06 | 4.04±0.21 |
| 137294 | 1999RE215 | 23.11±0.06 | 22.19±0.04 | 21.74±0.04 | — | — | 0.93±0.07 | 1.38±0.07 | 0.45±0.06 | — | 12.7±1.0 |
| 134860 | 2000OJ67 | — | — | — | 21.09±0.08 | 20.54±0.06 | — | — | — | 0.55±0.10 | — |
| 131695 | 2001XS254 | — | — | — | 21.80±0.06 | 21.20±0.06 | — | — | — | 0.60±0.08 | — |
| 130391 | 2000JG81 | 23.49±0.08 | 23.00±0.08 | 22.67±0.09 | — | — | 0.50±0.11 | 0.83±0.12 | 0.33±0.12 | — | -1.34±0.16 |
| 119070 | 2001KP77 | — | — | — | 21.38±0.07 | 20.86±0.04 | — | — | — | 0.53±0.08 | — |
| 118379 | 1999HC12 | 24.11±0.10 | — | 22.93±0.08 | 22.27±0.07 | 21.58±0.06 | — | 1.18±0.13 | — | 0.69±0.09 | 6.93±0.9 |
| 118228 | 1996TQ66 | 24.02±0.12 | 23.06±0.08 | 22.47±0.07 | — | — | 0.96±0.14 | 1.55±0.14 | 0.59±0.11 | — | 19.1±3.0 |
| 91133 | 1998HK151 | 22.52±0.04 | 21.76±0.03 | 21.47±0.04 | — | — | 0.76±0.05 | 1.05±0.06 | 0.30±0.05 | — | 3.54±0.18 |
| 87269 | 2000OO67 | 23.39±0.07 | 22.48±0.05 | 22.07±0.05 | — | — | 0.91±0.09 | 1.32±0.09 | 0.41±0.07 | — | 11.1±1.0 |
| 86177 | 1999RY215 | — | — | — | 21.69±0.06 | 21.01±0.05 | — | — | — | 0.68±0.08 | — |
| 86047 | 1999OY3 | — | — | — | 21.77±0.05 | 22.04±0.09 | — | — | — | -0.27±0.10 | — |
| 85633 | 1998KR65 | 23.63±0.08 | 22.85±0.07 | 22.29±0.06 | — | — | 0.77±0.11 | 1.34±0.10 | 0.57±0.09 | — | 11.7±1.2 |
| 85627 | 1998HP151 | 23.59±0.07 | — | 22.33±0.06 | — | — | — | 1.26±0.09 | — | — | 9.22±0.84 |
| 82075 | 2000YW134 | — | — | — | 19.66±0.04 | 19.10±0.03 | — | — | — | 0.56±0.05 | — |
| 80806 | 2000CM105 | — | — | — | 21.49±0.10 | 20.99±0.08 | — | — | — | 0.49±0.13 | — |
| 79360 | 1997CS29 | — | — | — | 20.30±0.07 | 19.74±0.06 | — | — | — | 0.57±0.09 | — |
| 69988 | 1998WA31 | 23.89±0.10 | 23.19±0.09 | 22.68±0.08 | — | — | 0.70±0.13 | 1.21±0.13 | 0.50±0.12 | — | 7.8±1.0 |
| 69987 | 1998WA25 | 24.02±0.08 | — | 22.92±0.07 | — | — | — | 1.10±0.11 | — | — | 4.8±0.5 |
| 66652 | BORASISI | — | — | — | 21.23±0.08 | 20.65±0.05 | — | — | — | 0.58±0.09 | — |
| 66452 | 1999OF4 | 23.65±0.09 | — | 22.45±0.07 | 21.66±0.09 | 21.13±0.07 | — | 1.19±0.11 | — | 0.53±0.11 | 7.49±0.88 |
| 60621 | 2000FE8 | — | — | — | 21.18±0.06 | 20.51±0.05 | — | — | — | 0.67±0.08 | — |
| 60620 | 2000FD8 | 23.06±0.07 | 22.20±0.05 | 21.64±0.05 | — | — | 0.86±0.09 | 1.42±0.09 | 0.56±0.07 | — | 14.4±1.3 |
| 60608 | 2000EE173 | 21.90±0.04 | 21.28±0.03 | 20.86±0.03 | — | — | 0.62±0.05 | 1.04±0.05 | 0.42±0.04 | — | 3.3±0.17 |
| 60454 | 2000CH105 | 23.42±0.09 | 22.67±0.05 | 22.15±0.07 | — | — | 0.75±0.10 | 1.27±0.11 | 0.52±0.09 | — | 9.5±1.1 |
| 58534 | 1997CQ29 | — | — | — | 21.83±0.07 | 21.27±0.04 | — | — | — | 0.56±0.08 | — |
| 54520 | 2000PJ30 | — | — | — | 22.93±0.11 | 22.34±0.13 | — | — | — | 0.59±0.17 | — |
| 48639 | 1995TL8 | — | — | — | 19.52±0.02 | 18.88±0.02 | — | — | — | 0.64±0.03 | — |
| 45802 | 2000PV29 | — | — | — | 22.92±0.09 | 22.36±0.12 | — | — | — | 0.56±0.15 | — |
| 45802 | 2000PV29 | — | — | — | 22.92±0.10 | 22.26±0.09 | — | — | — | 0.66±0.13 | — |
| 42301 | 2001UR163 | — | — | — | 19.45±0.02 | 18.83±0.01 | — | — | — | 0.62±0.03 | — |
| 38084 | 1999HB12 | 22.74±0.05 | 21.97±0.04 | 21.55±0.04 | — | — | 0.76±0.06 | 1.18±0.06 | 0.42±0.06 | — | 7.21±0.47 |
| 33340 | 1998VG44 | — | — | — | 19.85±0.02 | 19.31±0.02 | — | — | — | 0.54±0.03 | — |
| 33001 | 1997CU29 | 23.48±0.08 | 22.57±0.06 | 22.10±0.05 | 21.52±0.06 | 20.95±0.05 | 0.91±0.10 | 1.38±0.09 | 0.47±0.08 | 0.57±0.08 | 13.0±1.4 |
| 26375 | 1999DE9 | — | — | — | 19.14±0.03 | 18.64±0.02 | — | — | — | 0.50±0.04 | — |
| 24978 | 1998HJ151 | 23.95±0.08 | — | 22.83±0.07 | — | — | — | 1.12±0.11 | — | — | 5.31±0.55 |
| 24835 | 1995SM55 | — | — | — | 19.66±0.02 | 19.82±0.03 | — | — | — | -0.16±0.03 | — |

| Number | Object | F555W | F675W | F814W | F110W | F160W | F555W-F675W | F555W_F814W | F675W-F814W | F110W-F160W | $S^b$ |
|---|---|---|---|---|---|---|---|---|---|---|---|
| 20161 | 1996TR66 | — | — | — | 21.76±0.08 | 21.09±0.05 | — | — | — | 0.67±0.09 | — |
| 20108 | 1995QZ9 | 24.13±0.09 | — | 23.07±0.08 | — | — | — | 1.06±0.12 | — | — | 3.79±0.44 |
| 19521 | CHAOS | — | — | — | 19.61±0.02 | 19.06±0.02 | — | — | — | 0.55±0.03 | — |
| 19255 | 1994VK8 | 24.05±0.15 | 23.23±0.19 | 22.87±0.12 | 21.48±0.06 | 20.75±0.09 | 0.83±0.24 | 1.18±0.19 | 0.35±0.22 | 0.73±0.11 | 7.0±1.4 |
| 16684 | 1994JQ1 | 23.75±0.13 | 22.85±0.09 | 22.39±0.09 | — | — | 0.90±0.16 | 1.36±0.16 | 0.46±0.13 | — | 12.3±2.1 |
| 15883 | 1997CR29 | — | — | — | 22.30±0.10 | 21.60±0.07 | — | — | — | 0.70±0.12 | — |
| 15874 | 1996TL66 | — | — | — | 19.69±0.02 | 19.25±0.02 | — | — | — | 0.44±0.03 | — |
| 15836 | 1995DA2 | 23.78±0.07 | — | 22.69±0.06 | — | — | — | 1.10±0.09 | — | — | 4.54±0.41 |
| 15810 | 1994JR1 | 22.84±0.05 | 22.30±0.05 | 21.87±0.05 | 21.49±0.04 | 20.88±0.03 | 0.55±0.07 | 0.97±0.07 | 0.42±0.07 | 0.61±0.06 | 1.65±0.11 |
| 15809 | 1994JS | 23.71±0.13 | 22.78±0.09 | 22.44±0.09 | 21.95±0.55 | 21.29±0.06 | 0.92±0.16 | 1.26±0.16 | 0.34±0.13 | 0.66±0.55 | 9.5±1.6 |
| 15789 | 1993SC | 23.06±0.05 | 22.07±0.04 | 21.53±0.03 | 20.64±0.09 | 20.02±0.03 | 0.99±0.06 | 1.54±0.06 | 0.55±0.05 | 0.62±0.10 | 18.3±1.2 |
| 15788 | 1993SB | — | — | — | 21.47±0.04 | 20.83±0.04 | | — | — | 0.64±0.06 | — |
| 15760 | 1992QB1 | 23.69±0.06 | — | 22.50±0.05 | — | — | | 1.19±0.08 | — | — | 7.21±0.56 |
| — | 1995DB2 | 24.97±0.19 | — | 23.30±0.09 | — | — | — | 1.67±0.21 | — | — | 24.0±6.0 |
| — | 1995HM5 | 23.23±0.07 | 22.66±0.07 | 22.28±0.07 | — | — | 0.57±0.10 | 0.95±0.10 | 0.38±0.10 | — | 1.2±0.11 |
| — | 1996KV1 | — | — | — | 21.20±0.05 | 20.58±0.08 | — | — | — | 0.62±0.10 | — |
| — | 1996RQ20 | 23.33±0.07 | 22.59±0.06 | 22.17±0.05 | 21.47±0.05 | 20.94±0.04 | 0.74±0.09 | 1.17±0.09 | 0.42±0.08 | 0.53±0.07 | 6.38±0.58 |
| — | 1996RR20 | 24.04±0.19 | 23.13±0.08 | 22.57±0.08 | — | — | 0.91±0.21 | 1.47±0.21 | 0.56±0.11 | — | 16.1±4.0 |
| — | 1996TK66 | 23.18±0.06 | 22.37±0.05 | 22.05±0.05 | — | — | 0.81±0.08 | 1.14±0.08 | 0.32±0.07 | — | 5.58±0.44 |
| — | 1997CT29 | 23.78±0.10 | 22.78±0.07 | 22.44±0.07 | 21.89±0.05 | 21.31±0.05 | 1.01±0.12 | 1.34±0.12 | 0.34±0.10 | 0.58±0.07 | 11.7±1.5 |
| — | 1997CV29 | — | — | — | 21.88±0.08 | 21.38±0.07 | — | — | — | 0.50±0.11 | — |
| — | 1997RT5 | 23.90±0.09 | — | 22.75±0.08 | 21.90±0.06 | 21.38±0.06 | — | 1.15±0.12 | — | 0.51±0.09 | 6.11±0.72 |
| — | 1998FS144 | 23.50±0.07 | 22.63±0.05 | 22.25±0.06 | 21.79±0.05 | 21.25±0.05 | 0.87±0.09 | 1.25±0.09 | 0.39±0.08 | 0.54±0.07 | 8.93±0.81 |
| — | 1998KG62 | 23.48±0.08 | 22.55±0.06 | 22.11±0.05 | 21.85±0.05 | 21.28±0.06 | 0.93±0.10 | 1.37±0.09 | 0.43±0.08 | 0.57±0.08 | 12.7±1.3 |
| — | 1998KS65 | 23.92±0.08 | — | 22.71±0.06 | — | — | — | 1.21±0.10 | — | — | 7.77±0.81 |
| — | 1998KY61 | — | — | — | 22.55±0.11 | 22.06±0.13 | — | — | — | 0.50±0.17 | — |
| — | 1998UR43 | 24.00±0.13 | 23.15±0.11 | 22.80±0.10 | — | — | 0.85±0.17 | 1.20±0.16 | 0.35±0.15 | — | 7.5±1.3 |
| — | 1998UU43 | 23.29±0.08 | 22.53±0.07 | 22.26±0.07 | 21.58±0.05 | 20.89±0.04 | 0.75±0.11 | 1.03±0.11 | 0.27±0.10 | 0.69±0.07 | 3.06±0.32 |
| — | 1998WS31 | 23.81±0.10 | 23.00±0.08 | 22.50±0.07 | — | — | 0.81±0.13 | 1.31±0.12 | 0.50±0.11 | — | 10.7±1.4 |
| — | 1998WV24 | 23.51±0.08 | 22.86±0.06 | 22.53±0.07 | — | — | 0.65±0.10 | 0.99±0.11 | 0.34±0.09 | — | 1.88±0.2 |
| — | 1998WX24 | 23.60±0.09 | 22.64±0.06 | 22.28±0.06 | — | — | 0.95±0.11 | 1.32±0.11 | 0.37±0.08 | — | 11.1±1.3 |
| — | 1998WX31 | 22.82±0.05 | 21.97±0.04 | 21.46±0.03 | 21.21±0.05 | 20.66±0.04 | 0.85±0.06 | 1.36±0.06 | 0.51±0.05 | 0.55±0.06 | 12.3±0.8 |
| — | 1998WY24 | 23.24±0.07 | 22.44±0.05 | 21.91±0.05 | 21.61±0.05 | 21.05±0.06 | 0.81±0.09 | 1.33±0.09 | 0.52±0.07 | 0.56±0.08 | 11.4±1.04 |
| — | 1998WZ31 | 23.72±0.09 | 23.06±0.09 | 22.78±0.08 | 22.05±0.05 | 21.48±0.07 | 0.66±0.13 | 0.93±0.12 | 0.27±0.12 | 0.57±0.09 | 0.98±0.11 |
| — | 1998XY95 | 23.47±0.08 | 22.49±0.06 | 21.91±0.05 | — | — | 0.98±0.10 | 1.57±0.09 | 0.59±0.08 | — | 19.5±2.03 |
| — | 1999CD158 | — | — | — | 20.66±0.04 | 20.06±0.03 | — | — | — | 0.59±0.05 | — |
| — | 1999CF119 | 23.44±0.07 | 22.59±0.06 | 22.38±0.07 | 21.60±0.05 | 21.07±0.04 | 0.85±0.09 | 1.06±0.10 | 0.21±0.09 | 0.53±0.07 | 3.79±0.35 |
| — | 1999CH119 | — | — | — | 22.85±0.12 | 22.42±0.14 | — | — | — | 0.42±0.18 | — |
| — | 1999CJ119 | — | — | — | 21.71±0.05 | 21.06±0.04 | — | — | — | 0.65±0.06 | — |
| — | 1999CL119 | — | — | — | 21.69±0.07 | 21.04±0.07 | — | — | — | 0.65±0.11 | — |
| — | 1999CQ133 | — | — | — | 22.24±0.07 | 21.58±0.07 | — | — | — | 0.67±0.10 | — |
| — | 1999CX131 | 23.74±0.09 | 23.01±0.08 | 22.68±0.08 | 22.14±0.07 | 21.62±0.08 | 0.73±0.12 | 1.06±0.12 | 0.33±0.11 | 0.52±0.10 | 3.79±0.44 |
| — | 1999HJ12 | 24.10±0.11 | — | 22.76±0.08 | 22.29±0.43 | 21.60±0.08 | — | 1.34±0.14 | — | 0.69±0.44 | 11.7±1.7 |
| — | 1999OD4 | — | — | — | 22.11±0.06 | 21.45±0.07 | — | — | — | 0.66±0.10 | — |
| — | 1999OE4 | 23.82±0.13 | 23.16±0.18 | 22.89±0.13 | 21.52±0.05 | 21.01±0.04 | 0.66±0.22 | 0.92±0.18 | 0.27±0.22 | 0.50±0.07 | 0.76±0.13 |
| — | 1999OH4 | — | — | — | 22.57±0.16 | 22.11±0.19 | — | — | — | 0.46±0.25 | — |
| — | 1999OJ4 | — | — | — | 21.80±0.09 | 21.46±0.05 | — | — | — | 0.34±0.10 | — |
| — | 1999RC215 | 23.76±0.10 | 22.77±0.07 | 22.40±0.07 | 22.01±0.07 | 21.55±0.07 | 0.99±0.12 | 1.36±0.12 | 0.38±0.10 | 0.47±0.10 | 12.3±1.6 |

| Number | Object | F555W | F675W | F814W | F110W | F160W | F555W-F675W | F555W_ F814W | F675W-F814W | F110W-F160W | S[b] |
|---|---|---|---|---|---|---|---|---|---|---|---|
| — | 1999RX214 | 23.69±0.07 | — | 22.52±0.06 | — | — | — | 1.16±0.09 | — | — | 6.65±0.61 |
| — | 1999TR11 | 23.73±0.09 | 22.80±0.09 | 22.30±0.06 | — | — | 0.93±0.13 | 1.43±0.11 | 0.50±0.11 | — | 14.7±1.7 |
| — | 1999XY143 | 22.88±0.05 | 22.12±0.04 | 21.60±0.04 | — | — | 0.76±0.06 | 1.28±0.06 | 0.52±0.06 | — | 9.82±0.64 |
| — | 2000AF255 | — | — | — | 21.44±0.05 | 20.78±0.04 | — | — | — | 0.66±0.06 | — |
| — | 2000CE105 | 24.03±0.19 | — | 22.98±0.11 | 21.76±0.10 | 21.49±0.07 | — | 1.05±0.22 | — | 0.27±0.12 | 3.54±0.88 |
| — | 2000CF105 | — | — | — | 22.22±0.09 | 21.65±0.08 | — | — | — | 0.57±0.12 | — |
| — | 2000CG105 | 23.84±0.10 | 23.16±0.09 | 22.70±0.08 | 22.26±0.07 | 21.54±0.06 | 0.68±0.13 | 1.14±0.13 | 0.46±0.12 | 0.72±0.09 | 5.84±0.76 |
| — | 2000CK105 | 23.80±0.09 | 23.04±0.08 | 22.53±0.07 | 21.98±0.09 | 21.44±0.08 | 0.76±0.12 | 1.27±0.11 | 0.51±0.11 | 0.54±0.12 | 9.52±1.12 |
| — | 2000CL104 | 23.18±0.07 | 22.52±0.05 | 21.99±0.04 | 21.71±0.08 | 21.07±0.08 | 0.66±0.09 | 1.19±0.08 | 0.53±0.06 | 0.64±0.11 | 7.21±0.66 |
| — | 2000CO105 | 23.14±0.06 | 22.36±0.05 | 21.94±0.05 | 21.67±0.05 | 21.14±0.05 | 0.78±0.08 | 1.19±0.08 | 0.42±0.07 | 0.54±0.07 | 7.49±0.58 |
| — | 2000CP104 | 23.77±0.07 | — | 22.65±0.06 | 22.26±0.09 | 21.63±0.06 | — | 1.12±0.09 | — | 0.63±0.11 | 5.31±0.48 |
| — | 2000CQ105 | 23.64±0.09 | 22.88±0.07 | 22.65±0.08 | 22.15±0.06 | 21.55±0.07 | 0.76±0.11 | 0.99±0.12 | 0.23±0.11 | 0.60±0.09 | 2.11±0.25 |
| — | 2000CQ114 | 23.90±0.12 | 23.03±0.08 | 22.53±0.08 | 22.00±0.22 | 21.42±0.18 | 0.86±0.14 | 1.37±0.14 | 0.50±0.11 | 0.58±0.28 | 12.7±2.0 |
| — | 2000FS53 | 24.35±0.13 | — | 23.01±0.09 | — | — | — | 1.34±0.16 | — | — | 11.7±2.0 |
| — | 2000GV146 | 23.99±0.09 | — | 22.57±0.06 | — | — | — | 1.42±0.11 | — | — | 14.4±1.7 |
| — | 2000KK4 | 22.98±0.05 | 22.39±0.04 | 21.88±0.04 | — | — | 0.59±0.06 | 1.10±0.06 | 0.51±0.06 | — | 4.8±0.31 |
| — | 2000KL4 | 24.20±0.11 | — | 22.90±0.08 | — | — | — | 1.31±0.14 | — | — | 10.4±1.5 |
| — | 2000OU69 | — | — | — | 21.44±0.06 | 20.89±0.04 | — | — | — | 0.55±0.07 | — |
| — | 2000PD30 | 23.96±0.11 | — | 22.80±0.09 | 22.02±0.08 | 21.23±0.12 | — | 1.16±0.14 | — | 0.79±0.14 | 6.38±0.91 |
| — | 2000PE30 | — | — | — | 20.81±0.04 | 20.17±0.03 | — | — | — | 0.65±0.05 | — |
| — | 2000PH30 | — | — | — | 23.30±0.10 | 22.50±0.11 | — | — | — | 0.80±0.15 | — |
| — | 2001KD77 | — | — | — | 20.25±0.07 | 19.66±0.03 | — | — | — | 0.59±0.08 | — |
| — | 2001KY76 | — | — | — | 21.19±0.05 | 20.66±0.05 | — | — | — | 0.52±0.07 | — |
| — | 2001OG109 | — | — | — | 22.92±0.14 | 22.43±0.14 | — | — | — | 0.49±0.19 | — |
| — | 2001OK108 | — | — | — | 22.52±0.09 | 21.93±0.08 | — | — | — | 0.58±0.12 | — |
| — | 2001QC298 | — | — | — | 21.49±0.14 | 20.88±0.11 | — | — | — | 0.61±0.18 | — |
| — | 2001QX322 | — | — | — | 21.31±0.05 | 20.72±0.05 | — | — | — | 0.59±0.07 | — |
| — | 2001XU254 | — | — | — | 21.31±0.06 | 20.80±0.04 | — | — | — | 0.52±0.07 | — |

[a] Based on magnitude and color values from Cox, A. N. 2000, Allen's Astrophysical Quantities (New York: Springer), p. 341 and converted using *synphot*.
[b] Gradient between F555W and F814W.

**Table 3. Photometry (Johnson-Cousins-Bessel)**

| Number | Object | $V_{Johnson}$ | $R_{Cousins}$ | $I_{Cousins}$ | $J_{Bessel}$ | $H_{Bessel}$ | $V_J$-$R_C$ | $V_J$-$I_C$ | $R_C$-$I_C$ | $J_B$-$H_B$ |
|---|---|---|---|---|---|---|---|---|---|---|
| | Sun | -26.75 | — | — | — | — | 0.37±0.01 [a] | 0.69±0.01 [a] | 0.32±0.01 [a] | 0.31 [b] |
| 181902 | 1999RD215 | 23.61±0.07 | — | 22.73±0.07 | — | — | — | 0.88±0.10 | — | — |
| 181874 | 1999HW11 | 23.55±0.09 | 22.97±0.09 | 22.59±0.09 | — | — | 0.58±0.13 | 0.95±0.12 | 0.38±0.12 | — |
| 181871 | 1999CO153 | 23.92±0.08 | — | 22.49±0.05 | — | — | — | 1.43±0.10 | — | — |
| 181708 | 1993FW | 23.39±0.07 | 22.72±0.07 | 22.23±0.06 | 21.44±0.24 | 21.04±0.07 | 0.67±0.10 | 1.16±0.09 | 0.49±0.10 | 0.39±0.25 |
| 168703 | 2000GP183 | 22.01±0.04 | 21.54±0.05 | 21.11±0.04 | — | — | 0.46±0.06 | 0.90±0.06 | 0.43±0.06 | — |
| 168700 | 2000GE147 | 23.39±0.08 | 22.84±0.09 | 22.32±0.07 | — | — | 0.54±0.12 | 1.06±0.11 | 0.52±0.11 | — |
| 148209 | 2000CR105 | — | — | — | 22.43±0.28 | 22.02±0.10 | — | — | — | 0.41±0.30 |
| 138537 | 2000OK67 | 22.72±0.04 | 22.16±0.05 | 21.67±0.03 | 20.86±0.24 | 20.42±0.03 | 0.56±0.06 | 1.04±0.05 | 0.48±0.06 | 0.44±0.24 |
| 137294 | 1999RE215 | 23.09±0.06 | 22.35±0.05 | 21.75±0.04 | — | — | 0.74±0.08 | 1.34±0.08 | 0.59±0.06 | — |
| 134860 | 2000OJ67 | — | — | — | 20.90±0.24 | 20.50±0.07 | — | — | — | 0.39±0.25 |
| 131695 | 2001XS254 | — | — | — | 21.58±0.25 | 21.16±0.07 | — | — | — | 0.42±0.25 |
| 130391 | 2000JG81 | 23.47±0.08 | 23.10±0.10 | 22.67±0.09 | — | — | 0.37±0.13 | 0.80±0.12 | 0.43±0.14 | — |
| 119070 | 2001KP77 | — | — | — | 21.19±0.23 | 20.82±0.05 | — | — | — | 0.36±0.24 |
| 118379 | 1999HC12 | 24.09±0.10 | — | 22.94±0.08 | 22.02±0.29 | 21.53±0.07 | — | 1.15±0.13 | — | 0.48±0.29 |
| 118228 | 1996TQ66 | 24.00±0.13 | 23.23±0.10 | 22.49±0.08 | — | — | 0.77±0.16 | 1.51±0.15 | 0.75±0.13 | — |
| 91133 | 1998HK151 | 22.50±0.04 | 21.89±0.04 | 21.47±0.04 | — | — | 0.61±0.05 | 1.02±0.06 | 0.41±0.06 | — |
| 87269 | 2000OO67 | 23.37±0.07 | 22.63±0.06 | 22.08±0.05 | — | — | 0.74±0.09 | 1.29±0.09 | 0.55±0.08 | — |
| 86177 | 1999RY215 | — | — | — | 21.44±0.28 | 20.97±0.06 | — | — | — | 0.47±0.28 |
| 86047 | 1999OY3 | — | — | — | 21.63±0.24 | 22.01±0.09 | — | — | — | -0.38±0.26 |
| 85633 | 1998KR65 | 23.61±0.08 | 23.01±0.08 | 22.30±0.06 | — | — | 0.60±0.12 | 1.31±0.10 | 0.70±0.11 | — |
| 85627 | 1998HP151 | 23.57±0.07 | — | 22.34±0.06 | — | — | — | 1.23±0.10 | — | — |
| 82075 | 2000YW134 | — | — | — | 19.45±0.22 | 19.06±0.03 | — | — | — | 0.39±0.23 |
| 80806 | 2000CM105 | — | — | — | 21.32±0.24 | 20.96±0.09 | — | — | — | 0.37±0.25 |
| 79360 | 1997CS29 | — | — | — | 20.09±0.24 | 19.70±0.07 | — | — | — | 0.39±0.25 |
| 69988 | 1998WA31 | 23.87±0.10 | 23.34±0.11 | 22.69±0.09 | — | — | 0.53±0.15 | 1.18±0.13 | 0.65±0.14 | — |
| 69987 | 1998WA25 | 24.00±0.08 | — | 22.92±0.07 | — | — | — | 1.07±0.11 | — | — |
| 66652 | BORASISI | — | — | — | 20.15±0.24 | 19.78±0.06 | — | — | — | 0.36±0.24 |
| 66452 | 1999OF4 | 23.63±0.09 | — | 22.46±0.07 | 21.47±0.24 | 21.09±0.08 | — | 1.17±0.12 | — | 0.37±0.26 |
| 60621 | 2000FE8 | — | — | — | 20.94±0.27 | 20.47±0.06 | — | — | — | 0.47±0.28 |
| 60620 | 2000FD8 | 23.04±0.07 | 22.36±0.06 | 21.65±0.05 | — | — | 0.68±0.09 | 1.39±0.09 | 0.71±0.08 | — |
| 60608 | 2000EE173 | 21.88±0.04 | 21.41±0.04 | 20.86±0.03 | — | — | 0.47±0.05 | 1.01±0.05 | 0.54±0.05 | — |
| 60454 | 2000CH105 | 23.40±0.09 | 22.82±0.06 | 22.16±0.07 | — | — | 0.58±0.11 | 1.24±0.12 | 0.66±0.10 | — |

| Number | Object | $V_{Johnson}$ | $R_{Cousins}$ | $I_{Cousins}$ | $J_{Bessel}$ | $H_{Bessel}$ | $V_J-R_C$ | $V_J-I_C$ | $R_C-I_C$ | $J_B-H_B$ |
|--------|--------|---------------|---------------|---------------|--------------|--------------|-----------|-----------|-----------|-----------|
| 58534 | LOGOS | — | — | — | 21.62±0.25 | 21.23±0.05 | — | — | — | 0.39±0.25 |
| 54520 | 2000PJ30 | — | — | — | 22.71±0.27 | 22.30±0.14 | — | — | — | 0.41±0.30 |
| 48639 | 1995TL8 | — | — | — | 19.29±0.23 | 18.84±0.02 | — | — | — | 0.45±0.24 |
| 45802 | 2000PV29 | — | — | — | 22.72±0.24 | 22.32±0.13 | — | — | — | 0.39±0.28 |
| 45802 | 2000PV29 | — | — | — | 22.68±0.29 | 22.22±0.10 | — | — | — | 0.46±0.31 |
| 42301 | 2001UR163 | — | — | — | 19.22±0.23 | 18.79±0.01 | — | — | — | 0.43±0.23 |
| 38084 | 1999HB12 | 22.72±0.05 | 22.11±0.05 | 21.56±0.04 | — | — | 0.61±0.07 | 1.16±0.07 | 0.55±0.06 | — |
| 33340 | 1998VG44 | — | — | — | 19.66±0.21 | 19.27±0.02 | — | — | — | 0.38±0.21 |
| 33001 | 1997CU29 | 23.46±0.08 | 22.73±0.07 | 22.11±0.06 | 21.32±0.23 | 20.91±0.06 | 0.73±0.11 | 1.35±0.10 | 0.61±0.09 | 0.40±0.24 |
| 26375 | 1999DE9 | — | — | — | 18.96±0.20 | 18.60±0.02 | — | — | — | 0.35±0.20 |
| 24978 | 1998HJ151 | 23.93±0.08 | — | 22.83±0.07 | — | — | — | 1.09±0.11 | — | — |
| 24835 | 1995SM55 | — | — | — | 19.52±0.16 | 19.79±0.03 | — | — | — | -0.27±0.16 |
| 20161 | 1996TR66 | — | — | — | 21.52±0.29 | 21.05±0.06 | — | — | — | 0.47±0.29 |
| 20108 | 1995QZ9 | 24.11±0.09 | — | 23.07±0.09 | — | — | — | 1.03±0.12 | — | — |
| 19521 | CHAOS | — | — | — | 19.42±0.20 | 19.02±0.02 | — | — | — | 0.39±0.21 |
| 19255 | 1994VK8 | 24.03±0.15 | 23.37±0.22 | 22.88±0.13 | 21.22±0.28 | 20.70±0.10 | 0.66±0.26 | 1.15±0.20 | 0.49±0.25 | 0.51±0.30 |
| 16684 | 1994JQ1 | 23.73±0.13 | 23.01±0.11 | 22.40±0.10 | — | — | 0.72±0.17 | 1.33±0.17 | 0.60±0.15 | — |
| 15883 | 1997CR29 | — | — | — | 22.05±0.31 | 21.56±0.08 | — | — | — | 0.49±0.32 |
| 15874 | 1996TL66 | — | — | — | 19.53±0.17 | 19.22±0.02 | — | — | — | 0.32±0.17 |
| 15836 | 1995DA2 | 23.76±0.07 | — | 22.69±0.06 | — | — | — | 1.06±0.09 | — | — |
| 15810 | 1994JR1 | 22.82±0.05 | 22.42±0.06 | 21.87±0.05 | 21.27±0.24 | 20.84±0.03 | 0.40±0.08 | 0.94±0.07 | 0.55±0.08 | 0.43±0.24 |
| 15809 | 1994JS | 23.69±0.13 | 22.93±0.11 | 22.45±0.10 | 21.71±0.70 | 21.25±0.08 | 0.76±0.17 | 1.24±0.16 | 0.48±0.15 | 0.46±0.70 |
| 15789 | 1993SC | 23.04±0.05 | 22.24±0.05 | 21.55±0.03 | 20.41±0.28 | 19.98±0.04 | 0.80±0.07 | 1.49±0.06 | 0.69±0.06 | 0.43±0.28 |
| 15788 | 1993SB | — | — | — | 21.24±0.25 | 20.79±0.04 | — | — | — | 0.45±0.25 |
| 15760 | 1992QB1 | 23.67±0.06 | — | 22.51±0.05 | — | — | — | 1.16±0.08 | — | — |
| — | 1995DB2 | 24.95±0.20 | — | 23.32±0.10 | — | — | — | 1.63±0.22 | — | — |
| — | 1995HM5 | 23.21±0.07 | 22.77±0.09 | 22.28±0.07 | — | — | 0.43±0.11 | 0.93±0.10 | 0.49±0.11 | — |
| — | 1996KV1 | — | — | — | 20.97±0.24 | 20.54±0.09 | — | — | — | 0.43±0.26 |
| — | 1996RQ20 | 23.31±0.07 | 22.73±0.07 | 22.18±0.05 | 21.28±0.22 | 20.90±0.05 | 0.58±0.10 | 1.13±0.09 | 0.55±0.09 | 0.37±0.22 |
| — | 1996RR20 | 24.02±0.20 | 23.30±0.11 | 22.59±0.09 | — | — | 0.72±0.22 | 1.43±0.22 | 0.71±0.14 | — |
| — | 1996TK66 | 23.16±0.06 | 22.50±0.06 | 22.05±0.05 | — | — | 0.65±0.09 | 1.10±0.08 | 0.45±0.08 | — |
| — | 1997CT29 | 23.76±0.10 | 22.94±0.09 | 22.45±0.08 | 21.68±0.23 | 21.27±0.06 | 0.82±0.13 | 1.31±0.13 | 0.48±0.12 | 0.41±0.24 |
| — | 1997CV29 | — | — | — | 21.70±0.23 | 21.34±0.08 | — | — | — | 0.35±0.24 |
| — | 1997RT5 | 23.88±0.09 | — | 22.76±0.08 | 21.72±0.22 | 21.34±0.07 | — | 1.12±0.12 | — | 0.38±0.23 |

| Number | Object | V_Johnson | R_Cousins | I_Cousins | J_Bessel | H_Bessel | V_J-R_C | V_J-I_C | R_C-I_C | J_B-H_B |
|--------|--------|-----------|-----------|-----------|----------|----------|---------|---------|---------|---------|
| — | 1998FS144 | 23.48±0.07 | 22.77±0.06 | 22.26±0.06 | 21.60±0.22 | 21.21±0.06 | 0.70±0.09 | 1.22±0.10 | 0.52±0.09 | 0.38±0.23 |
| — | 1998KG62 | 23.46±0.08 | 22.71±0.07 | 22.12±0.06 | 21.65±0.22 | 21.24±0.07 | 0.75±0.11 | 1.34±0.10 | 0.58±0.09 | 0.40±0.23 |
| — | 1998KS65 | 23.90±0.08 | — | 22.72±0.06 | — | — | — | 1.18±0.10 | — | — |
| — | 1998KY61 | — | — | — | 22.37±0.24 | 22.02±0.14 | — | — | — | 0.34±0.28 |
| — | 1998UR43 | 23.98±0.13 | 23.29±0.13 | 22.81±0.11 | — | — | 0.69±0.19 | 1.17±0.17 | 0.48±0.17 | — |
| — | 1998UU43 | 23.27±0.08 | 22.66±0.09 | 22.26±0.07 | 21.33±0.27 | 20.84±0.04 | 0.61±0.12 | 1.00±0.11 | 0.39±0.11 | 0.48±0.28 |
| — | 1998WS31 | 23.79±0.10 | 23.15±0.10 | 22.51±0.08 | — | — | 0.64±0.14 | 1.28±0.13 | 0.64±0.12 | — |
| — | 1998WV24 | 23.49±0.08 | 22.98±0.08 | 22.53±0.07 | — | — | 0.51±0.11 | 0.95±0.11 | 0.45±0.11 | — |
| — | 1998WX24 | 23.58±0.09 | 22.79±0.07 | 22.29±0.06 | — | — | 0.79±0.12 | 1.29±0.11 | 0.50±0.10 | — |
| — | 1998WX31 | 22.80±0.05 | 22.13±0.05 | 21.47±0.03 | 21.02±0.23 | 20.62±0.05 | 0.67±0.07 | 1.33±0.06 | 0.65±0.06 | 0.39±0.23 |
| — | 1998WY24 | 23.22±0.07 | 22.60±0.06 | 21.92±0.05 | 21.41±0.23 | 21.01±0.07 | 0.62±0.09 | 1.30±0.09 | 0.67±0.08 | 0.39±0.24 |
| — | 1998WZ31 | 23.70±0.09 | 23.17±0.11 | 22.78±0.08 | 21.84±0.22 | 21.44±0.08 | 0.52±0.14 | 0.92±0.12 | 0.39±0.14 | 0.40±0.24 |
| — | 1998XY95 | 23.45±0.09 | 22.67±0.08 | 21.93±0.06 | — | — | 0.78±0.11 | 1.52±0.10 | 0.74±0.09 | — |
| — | 1999CD158 | — | — | — | 20.44±0.24 | 20.02±0.03 | — | — | — | 0.42±0.24 |
| — | 1999CF119 | 23.42±0.07 | 22.72±0.07 | 22.38±0.07 | 21.41±0.22 | 21.03±0.05 | 0.70±0.10 | 1.03±0.10 | 0.33±0.10 | 0.37±0.22 |
| — | 1999CH119 | — | — | — | 22.71±0.22 | 22.39±0.15 | — | — | — | 0.32±0.27 |
| — | 1999CJ119 | — | — | — | 21.47±0.26 | 21.02±0.05 | — | — | — | 0.45±0.26 |
| — | 1999CL119 | — | — | — | 21.45±0.27 | 21.00±0.08 | — | — | — | 0.45±0.28 |
| — | 1999CQ133 | — | — | — | 22.00±0.27 | 21.54±0.08 | — | — | — | 0.46±0.28 |
| — | 1999CX131 | 23.72±0.09 | 23.14±0.10 | 22.68±0.09 | 21.96±0.22 | 21.59±0.09 | 0.58±0.13 | 1.03±0.12 | 0.45±0.13 | 0.37±0.24 |
| — | 1999HJ12 | 24.08±0.11 | — | 22.77±0.09 | 22.04±0.59 | 21.56±0.10 | — | 1.31±0.14 | — | 0.48±0.60 |
| — | 1999OD4 | — | — | — | 21.87±0.27 | 21.41±0.08 | — | — | — | 0.46±0.28 |
| — | 1999OE4 | 23.80±0.13 | 23.27±0.21 | 22.89±0.14 | 21.34±0.21 | 20.98±0.05 | 0.52±0.24 | 0.91±0.19 | 0.38±0.25 | 0.36±0.22 |
| — | 1999OH4 | — | — | — | 22.41±0.26 | 22.08±0.20 | — | — | — | 0.34±0.33 |
| — | 1999OJ4 | — | — | — | 21.66±0.19 | 21.43±0.05 | — | — | — | 0.23±0.20 |
| — | 1999RC215 | 23.74±0.10 | 22.93±0.09 | 22.41±0.08 | 21.84±0.21 | 21.52±0.08 | 0.81±0.13 | 1.33±0.13 | 0.51±0.12 | 0.33±0.22 |
| — | 1999RX214 | 23.67±0.07 | — | 22.53±0.06 | — | — | — | 1.14±0.09 | — | — |
| — | 1999TR11 | 23.71±0.09 | 22.96±0.10 | 22.31±0.07 | — | — | 0.75±0.14 | 1.40±0.11 | 0.65±0.12 | — |
| — | 1999XY143 | 22.86±0.05 | 22.27±0.05 | 21.61±0.04 | — | — | 0.59±0.07 | 1.25±0.07 | 0.66±0.06 | — |
| — | 2000AF255 | — | — | — | 21.20±0.26 | 20.74±0.05 | — | — | — | 0.46±0.27 |
| — | 2000CE105 | 24.01±0.19 | — | 22.98±0.12 | 21.62±0.17 | 21.46±0.07 | — | 1.02±0.22 | — | 0.16±0.19 |
| — | 2000CF105 | — | — | — | 22.01±0.25 | 21.61±0.09 | — | — | — | 0.40±0.27 |
| — | 2000CG105 | 23.82±0.10 | 23.29±0.11 | 22.70±0.09 | 22.00±0.29 | 21.49±0.07 | 0.52±0.15 | 1.11±0.13 | 0.59±0.14 | 0.50±0.30 |
| — | 2000CK105 | 23.78±0.09 | 23.19±0.09 | 22.54±0.07 | 21.79±0.24 | 21.40±0.09 | 0.59±0.13 | 1.24±0.12 | 0.65±0.12 | 0.38±0.26 |

| Number | Object | $V_{Johnson}$ | $R_{Cousins}$ | $I_{Cousins}$ | $J_{Bessel}$ | $H_{Bessel}$ | $V_J-R_C$ | $V_J-I_C$ | $R_C-I_C$ | $J_B-H_B$ |
|---|---|---|---|---|---|---|---|---|---|---|
| — | 2000CL104 | 23.16±0.07 | 22.66±0.06 | 22.00±0.04 | 21.48±0.27 | 21.03±0.09 | 0.50±0.09 | 1.16±0.08 | 0.66±0.08 | 0.45±0.28 |
| — | 2000CO105 | 23.12±0.06 | 22.50±0.06 | 21.95±0.05 | 21.48±0.22 | 21.10±0.06 | 0.62±0.08 | 1.17±0.08 | 0.55±0.08 | 0.37±0.22 |
| — | 2000CP104 | 23.75±0.07 | — | 22.65±0.06 | 22.03±0.28 | 21.59±0.07 | — | 1.09±0.09 | — | 0.44±0.29 |
| — | 2000CQ105 | 23.62±0.09 | 23.00±0.09 | 22.65±0.09 | 21.93±0.25 | 21.51±0.08 | 0.62±0.13 | 0.96±0.12 | 0.35±0.12 | 0.42±0.26 |
| — | 2000CQ114 | 23.88±0.12 | 23.19±0.10 | 22.54±0.09 | 21.80±0.35 | 21.38±0.20 | 0.69±0.16 | 1.34±0.15 | 0.64±0.13 | 0.41±0.40 |
| — | 2000FS53 | 24.33±0.13 | — | 23.02±0.10 | — | — | — | 1.31±0.17 | — | — |
| — | 2000GV146 | 23.97±0.09 | — | 22.58±0.07 | — | — | — | 1.39±0.11 | — | — |
| — | 2000KK4 | 22.96±0.05 | 22.52±0.05 | 21.88±0.04 | — | — | 0.43±0.07 | 1.07±0.07 | 0.64±0.07 | — |
| — | 2000KL4 | 24.18±0.11 | — | 22.91±0.09 | — | — | — | 1.27±0.14 | — | — |
| — | 2000OU69 | — | — | — | 21.25±0.23 | 20.85±0.05 | — | — | — | 0.39±0.24 |
| — | 2000PD30 | 23.94±0.11 | — | 22.81±0.10 | 21.74±0.31 | 21.18±0.13 | — | 1.13±0.15 | — | 0.56±0.34 |
| — | 2000PE30 | — | — | — | 20.57±0.25 | 20.13±0.03 | — | — | — | 0.44±0.25 |
| — | 2000PH30 | — | — | — | 23.00±0.34 | 22.45±0.12 | — | — | — | 0.56±0.36 |
| — | 2001KD77 | — | — | — | 20.03±0.26 | 19.62±0.04 | — | — | — | 0.41±0.26 |
| — | 2001KY76 | — | — | — | 21.01±0.22 | 20.62±0.06 | — | — | — | 0.39±0.22 |
| — | 2001OG109 | — | — | — | 22.75±0.26 | 22.40±0.15 | — | — | — | 0.36±0.30 |
| — | 2001OK108 | — | — | — | 22.32±0.26 | 21.89±0.09 | — | — | — | 0.42±0.28 |
| — | 2001QC298 | — | — | — | 21.27±0.30 | 20.84±0.12 | — | — | — | 0.43±0.33 |
| — | 2001QX322 | — | — | — | 21.09±0.24 | 20.68±0.06 | — | — | — | 0.41±0.24 |
| — | 2001XU254 | — | — | — | 21.13±0.22 | 20.76±0.05 | — | — | — | 0.36±0.23 |

[a] Values converted from Cox, A. N. 2000, Allen's Astrophysical Quantities (New York: Springer), p. 341 using the transformation equations of Fernie, 1983. The values recorded in Cox are for Johnson filters, the values given here are for Cousins filters.
[b] Value from: Cox, A. N. 2000, Allen's Astrophysical Quantities (New York: Springer), p. 341.

**Table 4. H$_v$ magnitudes**

| Number | Object | HST | MPC | Tegler |
|--------|--------|-----|-----|--------|
| 181902 | 1999RD215 | 7.55±0.07 | 7.4 | — |
| 181874 | 1999HW11 | 7.16±0.09 | 6.8 | — |
| 181871 | 1999CO153 | 7.60±0.08 | 7.4 | — |
| 181708 | 1993FW | 7.05±0.07 | 7.0 | 7.09±0.01[e] |
| 168703 | 2000GP183 | 6.06±0.04 | 6.6 | — |
| 168700 | 2000GE147 | 8.28±0.08 | 8.5 | — |
| 138537 | 2000OK67 | 6.41±0.04 | 5.9 | — |
| 137294 | 1999RE215 | 6.77±0.06 | 6.7 | — |
| 130391 | 2000JG81 | 7.87±0.08 | 8.0 | — |
| 118379 | 1999HC12 | 7.89±0.10 | 7.6 | — |
| 118228 | 1996TQ66 | 8.36±0.13 | 7.1 | 7.69±0.11[a] |
| 91133 | 1998HK151 | 7.37±0.04 | 7.6 | — |
| 87269 | 2000OO67 | 9.75±0.07 | 9.2 | 9.82±0.12[d] |
| 85633 | 1998KR65 | 7.08±0.08 | 6.7 | 7.10±0.03[e] |
| 85627 | 1998HP151 | 7.08±0.07 | 7.4 | — |
| 69988 | 1998WA31 | 7.72±0.10 | 7.5 | — |
| 69987 | 1998WA25 | 7.60±0.08 | 7.2 | — |
| 66452 | 1999OF4 | 6.90±0.09 | 6.9 | — |
| 60620 | 2000FD8 | 6.82±0.07 | 6.6 | — |
| 60608 | 2000EE173 | 8.31±0.04 | 8.6 | 8.49±0.01[d] |
| 60454 | 2000CH105 | 6.87±0.09 | 6.3 | — |
| 38084 | 1999HB12 | 7.07±0.05 | 7.4 | 7.04±0.01[d] |
| 33001 | 1997CU29 | 6.78±0.08 | 6.6 | 6.68±0.08[e] |
| 24978 | 1998HJ151 | 7.49±0.08 | 7.5 | 7.67±0.02[e] |
| 20108 | 1995QZ9 | 8.54±0.09 | 7.9 | 8.58±0.02[e] |
| 19255 | 1994VK8 | 7.46±0.15 | 7.0 | 7.53±0.02[e] |
| 16684 | 1994JQ1 | 7.22±0.13 | 6.9 | 7.14±0.03[e] |
| 15836 | 1995DA2 | 8.37±0.07 | 8.1 | — |
| 15810 | 1994JR1 | 7.25±0.05 | 7.7 | 7.35±0.10[a] |
| 15809 | 1994JS | 8.04±0.13 | 7.8 | — |

| Number | Object | HST | MPC | Tegler |
|--------|--------|-----|-----|--------|
| 15789 | 1993SC | 7.39±0.05 | 7.0 | 7.30±0.06[b] |
| 15760 | 1992QB1 | 7.38±0.06 | 7.2 | 7.61±0.03[c] |
| — | 1995DB2 | 8.81±0.20 | 7.6 | — |
| — | 1995HM5 | 7.95±0.07 | 8.3 | 8.29±0.10[a] |
| — | 1996RQ20 | 7.26±0.07 | 7.0 | 7.00±0.07[a] |
| — | 1996RR20 | 7.47±0.20 | 6.8 | 7.20±0.01[c] |
| — | 1996TK66 | 6.69±0.06 | 6.4 | 6.75±0.03[c] |
| — | 1997CT29 | 7.06±0.10 | 6.6 | 7.19±0.11[c] |
| — | 1997RT5 | 7.41±0.09 | 7.3 | — |
| — | 1998FS144 | 7.18±0.07 | 6.7 | — |
| — | 1998KG62 | 6.83±0.08 | 6.6 | — |
| — | 1998KS65 | 7.56±0.08 | 7.7 | 7.63±0.02[c] |
| — | 1998UR43 | 8.72±0.13 | 8.3 | — |
| — | 1998UU43 | 7.32±0.08 | 7.2 | — |
| — | 1998WS31 | 8.57±0.10 | 8.3 | — |
| — | 1998WV24 | 7.44±0.08 | 7.5 | 7.43±0.01[c] |
| — | 1998WX24 | 6.85±0.09 | 6.6 | 6.79±0.04[c] |
| — | 1998WX31 | 6.74±0.05 | 6.6 | — |
| — | 1998WY24 | 7.01±0.07 | 6.7 | — |
| — | 1998WZ31 | 8.24±0.09 | 8.1 | — |
| — | 1998XY95 | 6.99±0.09 | 6.2 | — |
| — | 1999CF119 | 7.37±0.07 | 7.3 | 7.42±0.04[d] |
| — | 1999CX131 | 7.36±0.09 | 7.0 | — |
| — | 1999HJ12 | 7.44±0.11 | 7.4 | — |
| — | 1999OE4 | 7.23±0.13 | 7.0 | — |
| — | 1999RC215 | 7.19±0.10 | 6.9 | — |
| — | 1999RX214 | 6.87±0.07 | 6.8 | — |
| — | 1999TR11 | 8.70±0.09 | 8.4 | 8.63±0.07[c] |
| — | 1999XY143 | 6.45±0.05 | 6.0 | — |
| — | 2000CE105 | 7.64±0.19 | 6.8 | — |
| — | 2000CG105 | 7.12±0.10 | 6.5 | — |
| — | 2000CK105 | 6.87±0.09 | 6.2 | — |

| Number | Object | HST | MPC | Tegler |
|---|---|---|---|---|
| — | 2000CL104 | 6.77±0.07 | 6.3 | — |
| — | 2000CO105 | 6.21±0.06 | 6.0 | — |
| — | 2000CP104 | 7.07±0.07 | 6.7 | — |
| — | 2000CQ105 | 6.46±0.09 | 6.0 | 6.29±0.01[d] |
| — | 2000CQ114 | 7.20±0.12 | 7.0 | — |
| — | 2000FS53 | 7.87±0.13 | 7.7 | 7.88±0.06[e] |
| — | 2000GV146 | 7.55±0.09 | 7.6 | — |
| — | 2000KK4 | 6.34±0.05 | 6.1 | 6.46±0.02[e] |
| — | 2000KL4 | 7.93±0.11 | 7.7 | — |
| — | 2000PD30 | 7.16±0.11 | 7.3 | — |

[a] Tegler, S. C., & Romanishin, W. 1998, Nature, 392, 49.

[b] Romanishin, W., & Tegler, S. C. 1999, Nature, 398, 129.

[c] Tegler, S. C., & Romanishin, W. 2000, Nature, 407, 979.

[d] Tegler, S. C., Romanishin, W., & Consolmagno, G. J. 2003a, ApJ, 599, 49.

[e] Tegler, S. C., & Romanishin, W. 2003b, Icarus, 161, 181.

**Table 5. Spearman's ρ rank correlation**

| Sample[a] | HST | | | Literature+HST | | |
|---|---|---|---|---|---|---|
| | **Nobj** | **ρ** | **Sig**[b] | **Nobj** | **ρ** | **Sig**[b] |
| **Hot-Cen (V-I) inc** | — | — | — | 90 | -0.430 | **2.4E-05** |
| **Hot (V-I) inc** | 20 | -0.150 | 0.527 | 129 | -0.303 | **4.9E-04** |
| Hot (V-I) peri | 20 | -0.025 | 0.917 | 129 | 0.222 | 0.011 |
| Hot-Cen (V-I) peri | — | — | — | 90 | 0.253 | 0.016 |
| Hot-Cen (V-I) ecc | — | — | — | 90 | -0.239 | 0.024 |
| ***Hot-Cen (V-I) hmag*** | — | — | — | 90 | 0.198 | ***0.061*** |
| Res (V-I) semi | 22 | -0.030 | 0.893 | 67 | 0.223 | 0.069 |
| Res (V-I) peri | 22 | -0.071 | 0.753 | 67 | 0.190 | 0.123 |
| Cold (V-I) peri | 31 | -0.102 | 0.584 | 57 | 0.194 | 0.149 |
| Hot (V-I) semi | 20 | -0.171 | 0.472 | 129 | 0.127 | 0.153 |
| Res (V-I) inc | 22 | -0.225 | 0.314 | 67 | -0.173 | 0.161 |
| Hot (V-I) ecc | 20 | -0.210 | 0.373 | 129 | -0.116 | 0.192 |
| Cen (V-I) semi | — | — | — | 39 | 0.184 | 0.263 |
| Hot-Cen (V-I) semi | — | — | — | 90 | -0.105 | 0.324 |
| Cen (V-I) ecc | — | — | — | 39 | 0.137 | 0.405 |
| Cen (V-I) inc | — | — | — | 39 | -0.133 | 0.419 |
| Res (V-I) ecc | 22 | -0.332 | 0.131 | 67 | 0.082 | 0.508 |
| Cen (V-I) hmag | — | — | — | 39 | -0.104 | 0.529 |
| Cold (V-I) hmag | 31 | -0.008 | 0.965 | 57 | 0.073 | 0.588 |
| ***Res (V-I) hmag*** | 22 | -0.429 | 0.047 | 67 | -0.048 | ***0.701*** |
| Cen (V-I) peri | — | — | — | 39 | -0.057 | 0.730 |
| Cold (V-I) ecc | 31 | 0.223 | 0.228 | 57 | -0.046 | 0.733 |
| Hot (V-I) hmag | 20 | -0.027 | 0.910 | 129 | 0.029 | 0.748 |
| Cold (V-I) semi | 31 | 0.155 | 0.405 | 57 | 0.027 | 0.839 |
| Cold (V-I) inc | 31 | 0.125 | 0.504 | 57 | 0.026 | 0.847 |
| ***Res (J-H) hmag*** | 18 | 0.214 | 0.393 | 32 | 0.786 | ***1.0E-07*** |
| ***Hot-Cen (J-H) hmag*** | — | — | — | 58 | 0.547 | ***9.0E-06*** |
| Cen (J-H) hmag | — | — | — | 26 | 0.614 | 0.001 |
| Hot (J-H) semi | 31 | 0.260 | 0.158 | 84 | 0.238 | 0.029 |

| Sample[a] | HST | | | Literature+HST | | |
|---|---|---|---|---|---|---|
| | Nobj | ρ | Sig[b] | Nobj | ρ | Sig[b] |
| Hot (J-H) hmag | 31 | 0.103 | 0.580 | 84 | 0.230 | 0.036 |
| **Hot-Cen (J-H) inc** | — | — | — | 58 | -0.267 | **0.043** |
| **Hot (J-H) inc** | 31 | -0.156 | 0.403 | 84 | -0.219 | **0.045** |
| Hot-Cen (J-H) semi | — | — | — | 58 | 0.235 | 0.076 |
| Cen (J-H) peri | — | — | — | 26 | 0.340 | 0.089 |
| Res (J-H) semi | 18 | -0.232 | 0.354 | 32 | 0.250 | 0.167 |
| Hot (J-H) peri | 31 | -0.098 | 0.601 | 84 | 0.132 | 0.232 |
| Cen (J-H) ecc | — | — | — | 26 | -0.216 | 0.289 |
| Cen (J-H) semi | — | — | — | 26 | 0.172 | 0.402 |
| Res (J-H) peri | 18 | -0.223 | 0.373 | 32 | 0.149 | 0.416 |
| Cold (J-H) ecc | 31 | 0.034 | 0.855 | 31 | 0.150 | 0.422 |
| Cold (J-H) semi | 31 | 0.160 | 0.389 | 31 | 0.131 | 0.483 |
| Cold (J-H) inc | 31 | 0.184 | 0.322 | 31 | 0.124 | 0.507 |
| Hot-Cen (J-H) peri | — | — | — | 58 | 0.087 | 0.515 |
| Cen (J-H) inc | — | — | — | 26 | 0.100 | 0.628 |
| Cold (J-H) peri | 31 | 0.053 | 0.775 | 31 | -0.074 | 0.692 |
| Cold (J-H) hmag | 31 | -0.037 | 0.843 | 31 | -0.056 | 0.764 |
| Res (J-H) ecc | 18 | -0.193 | 0.442 | 32 | 0.027 | 0.885 |
| Hot-Cen (J-H) ecc | — | — | — | 58 | 0.017 | 0.899 |
| Res (J-H) inc | 18 | 0.071 | 0.778 | 32 | 0.011 | 0.951 |
| Hot (J-H) ecc | 31 | 0.137 | 0.464 | 84 | 0.003 | 0.981 |

**Bold** entries are correlations of interest in the optical, and ***Bold-italic*** are correlations of interest in the infrared.

[a] "Hot" is the sample as defined in the text. "Hot-Cen" is the Hot sample as defined in the text with the Centaurs excluded. We do not run the calculations excluding Centaurs for the HST only dataset as there are only 3 objects that fall into the Centaur classification.

[b] The significance is a value in the interval [0.0, 1.0] where a small value indicates a significant correlation. A 3-sigma result yields a significance of ~0.001.

**Table 6. Kolmogorov-Smirnov Test Results**

| Color | Sample1[b] | Sample2[b] | D | %[a] | N1 | N2 |
|-------|-----------|-----------|------|--------|-----|-----|
| F555W-F814W | COLD | HOT | 0.35 | 92.42 | 31 | 20 |
| F555W-F814W | COLD | RES | 0.29 | 81.92 | 31 | 22 |
| F555W-F814W | HOT | RES | 0.22 | 18.07 | 20 | 22 |
| F110W-F160W | COLD | HOT | 0.29 | 88.00 | 31 | 31 |
| F110W-F160W | COLD | RES | 0.30 | 77.91 | 31 | 18 |
| F110W-F160W | HOT | RES | 0.17 | 13.63 | 31 | 18 |
| V-I | COLD | HOT | 0.49 | >99.99 | 53 | 126 |
| V-I | COLD | HOT-CEN | 0.49 | >99.99 | 53 | 87 |
| V-I | COLD | CEN | 0.53 | >99.99 | 53 | 39 |
| V-I | COLD | RES | 0.41 | 99.99 | 53 | 67 |
| V-I | RES | CEN | 0.29 | 97.60 | 67 | 39 |
| V-I | HOT | CEN | 0.22 | 86.99 | 87 | 39 |
| V-I | HOT | RES | 0.16 | 83.58 | 126 | 67 |
| V-I | HOT-CEN | RES | 0.14 | 62.34 | 87 | 67 |
| J-H | COLD | CEN | 0.48 | 99.86 | 32 | 26 |
| J-H | COLD | HOT | 0.36 | 99.72 | 32 | 85 |
| J-H | COLD | HOT-CEN | 0.31 | 97.36 | 32 | 59 |
| J-H | HOT-CEN | RES | 0.11 | 94.85 | 59 | 33 |
| J-H | RES | CEN | 0.30 | 87.24 | 33 | 26 |
| J-H | COLD | RES | 0.24 | 73.10 | 32 | 33 |
| J-H | HOT | CEN | 0.21 | 64.86 | 59 | 26 |
| J-H | HOT | RES | 0.16 | 49.55 | 85 | 33 |

[a] The level of confidence that the two groups are not drawn from the same parent population.

[b] "Hot" is the sample as defined in the text. "Hot-Cen" is the Hot sample as defined in the text with the Centaurs excluded. We do not run the calculations excluding Centaurs for the HST only dataset as there are only 3 objects that fall into the Centaur classification.